\documentclass[draftclsnofoot,onecolumn,12pt]{IEEEtran}

\usepackage{verbatim}
\usepackage{amsfonts}
\usepackage{amssymb}
\usepackage{stfloats}
\usepackage{graphicx}
\usepackage{psfrag}
\usepackage{amsmath}
\usepackage{array}
\usepackage{epstopdf}
\usepackage{authblk}
\usepackage{graphicx} 
\usepackage{amsthm} 
\usepackage{lipsum}
\usepackage{verbatim} 
\usepackage{authblk}
\usepackage{mathtools}
\usepackage{cuted}
\usepackage[lined,boxed,ruled]{algorithm2e}
\usepackage{booktabs}
\usepackage{bm}
\usepackage[hidelinks]{hyperref}

\ifCLASSOPTIONcompsoc
\usepackage[caption=false,font=normalsize,labelfon
t=sf,textfont=sf]{subfig}
\else
\usepackage[caption=false,font=footnotesize]{subfig}
\fi

\usepackage[usenames]{color}

\newtheorem{lemma}{Lemma}

\def\proof{\noindent{\emph{Proof:} }}

\def\phi{\varphi}

\def\({\left(}
\def\){\right)}

\def\b0{{\mathbf{0}}}

\newcommand{\diag}{\mathrm{diag}}

\makeatletter
\newcommand{\removelatexerror}{\let\@latex@error\@gobble}
\makeatother

\begin{document}

\graphicspath{{figure/}}

\title{\huge Task-Oriented Over-the-Air Computation for Multi-Device Edge AI }
\author{Dingzhu Wen, Xiang Jiao, Peixi Liu, Guangxu Zhu, Yuanming Shi, and Kaibin Huang \thanks{\setlength{\baselineskip}{13pt} \noindent D. Wen is with Network Intelligence Center, School of Information Science and Technology, ShanghaiTech University, Shanghai, China (e-mail: wendzh@shanghaitech.edu.cn), and was with Shenzhen Research Institute of Big Data, Shenzhen, China. (Corresponding author: G. Zhu)

Xiang Jiao and P. Liu are with State Key Laboratory of Advanced Optical Communication Systems and Networks, School of Electronics, Peking University, China, and Shenzhen Research Institute of Big Data, Shenzhen, China (e-mail:  jiaoxiang@stu.pku.edu.cn, liupeixi@pku.edu.cn).

G. Zhu is with Shenzhen Research Institute of Big Data, Shenzhen, China (e-mail: gxzhu@sribd.cn). 

Y. Shi is  with Network Intelligence Center, School of Information Science and Technology, ShanghaiTech University, Shanghai, China (e-mail: shiym@shanghaitech.edu.cn). 

K. Huang is with Departement of Electrical and Electronic Engineering, The  University of  Hong Kong, Hong Kong (e-mail:huangkb@eee.hku.hk).

} }

\maketitle

\begin{abstract}
Edge inference refers to the use of artificial intelligent (AI) models at the network edge to provide mobile devices inference services and thereby enable intelligent services such as auto-driving and Metaverse towards 6G. However, departing from the classic paradigm of data-centric designs, the 6G networks for supporting edge AI features task-oriented techniques that focus on effective and efficient execution of AI task. Targeting end-to-end system performance, such techniques are sophisticated as they aim to seamlessly integrate sensing (data acquisition), communication (data transmission), and computation (data processing). Aligned with the paradigm shift, a task-oriented over-the-air computation (AirComp) scheme is proposed in this paper for multi-device split-inference system. In the considered system, local feature vectors, which are extracted from the real-time noisy sensory data on devices, are aggregated over-the-air by exploiting the waveform superposition in a multiuser channel. Then the aggregated features as received at a server are fed into an inference model with the result used for decision making or control of actuators. To design inference-oriented AirComp, the transmit precoders at edge devices and receive beamforming at edge server are jointly optimized to rein in the aggregation error and maximize the inference accuracy. The problem is made tractable by measuring the inference accuracy using a surrogate metric called discriminant gain, which measures the discernibility of two object classes in the application of object/event classification. It is discovered that the conventional AirComp beamforming design for minimizing the mean square error in generic AirComp with respect to the noiseless case may not lead to the optimal classification accuracy. The reason is due to the overlooking of the fact that feature dimensions have different sensitivity towards aggregation errors and are thus of different importance levels for classification. This issue is addressed in this work via a new task-oriented AirComp scheme designed by directly maximizing the derived discriminant gain. However, the resultant problem of joint transmit precoding and receive beamforming is nonconvex and difficult to solve due to the complicated form of discriminant gain and the coupling between the control variables. We overcome the difficulty using the successive convex approximation. The performance gain of the proposed task-oriented scheme over the conventional schemes is verified by extensive experiments targeting the application of human motion recognition.

\end{abstract}

\section{Introduction}
One main function of 6G networks is to provide artificial intelligent (AI) services such as auto-driving, eHealth, and Metaverse, at the network edge \cite{letaief2019the, letaief2021edge, zhu2020toward, shi2020communication,wen2022federated}. Existing data-driven services, i.e., transmitting multi-media data like voice, text, image, and video, focus on throughput maximization where only the communication process is considered. On the contrary, AI services at the network edge are goal oriented and aim to achieve the required accuracy and latency for completing a specific task (see, e.g., \cite{shi2022mobile,wen2020joint,yang2020energy}). The task execution typically involves the tight integration of three processes, i.e.,  \emph{sensing} for real-time data acquisition, \emph{communication} for data transmission, and \emph{computation} for decision making \cite{liu2022vertical,wen2022task}. 
Then to efficiently support edge-intelligence services drives an ongoing paradigm shift in wireless technologies from the traditional data-centric design toward the task-oriented design for 6G \cite{strinati20216g, ma2021frac, lan2021semantic, xie2021task}. On the other hand, edge-intelligence services rely on the deployment of trained AI models at the network edge for decision making and timely response to a dynamic environment  \cite{zhou2019edge, soifer2019deep, jang2021microservice}. This gives rise to an emerging research area, called edge inference. The design of edge inference faces two main challenges. On one hand, the task-oriented techniques for efficient edge inference must integrate sensing, communication, and computation, and thus their designs are sophisticated and cross-disciplinary. On the other hand, efficient edge inference has to overcome a communication bottleneck caused by the low-latency requirements of real-time AI services (e.g., human motion recognition in auto-driving) and the need to upload the sensory data from potentially many devices for aggregation to suppress sensing noise. One promising solution for these challenges is called over-the-air computation (AirComp) that leverages the waveform-superposition property of a multi-access channel to realize over-the-air aggregation of analog modulated sensory data simultaneously transmitted by multiple devices. The communication-and-computing integration and the enabled simultaneous access promise to dramatically reduce multi-access latency and suppress communication overhead when there are many devices. In this work, we design task-oriented AirComp techniques to support communication-efficient edge inference.


Recent years have witnessed the advancements of edge inference on different fronts. Split inference is arguably the most popular edge-inference architecture, which divides an AI model into two parts: one deployed on resource-limited devices for feature extraction [via, e.g., principal component analysis (PCA) or using a convolutional neural network], and the other at an edge server for completing the remaining computation-intensive inference task. Thereby, the avoidance of direct data uploading helps preserve data privacy and the offloading of computation-intensive task to the server overcomes the devices' resource constraints. These advantages motivate us to adopt split inference in this work. One main research focus on edge split inference is to balance the trade-off between the computation and communication overhead on edge device via, e.g., compressing the feature map of the split layer \cite{shao2020communication,huang2020dynamic,shao2021learning}, a two-step pruning strategy \cite{shi2020device}, progressive feature transmission \cite{lan2021progressive}, setting early existing point \cite{li2019edge,liu2022resource}, and joint source and channel coding using deep neural networks \cite{jankowski2020joint}. However, the aforementioned designs focus on the case of a single edge device. This pertains to scenarios where the device either senses the source in a narrow view to obtain highly accurate sensory data by focusing a single angle, or obtains a noise-corrupted wide-view sensory data for wide angle object detection by, e.g., scanning from angle to angle \cite{yi2019wide}. To address the incomplete feature space caused by the narrow-view sensing, a multi-device cooperated multi-view edge inference scheme is proposed in \cite{wen2022task} to maximize the inference accuracy via the design of task-oriented sensing, computation, and communication integration. However, the issue of suppressing the noise of wide-view sensory data for inference accuracy enhancement remains unresolved and is the theme of this paper. 

In this work, a multi-device edge inference system is considered. Each device obtains a noise-corrupted version of the ground-true wide-view sensory data and extracts from it a noisy local feature vector using simple linear operations like PCA. To suppress the sensing noise, we adopt a common approach, which averages out the noise via a weighted sum of all local feature vectors \cite{luo2005universal,xiao2006power}. To this end, the technique of AirComp can be employed to enhance the communication efficiency due to its capability in supporting fast data aggregation from a large number of devices \cite{zhu2021over}. Specifically, in AirComp, signals from all devices are allowed to transmit simultaneously over the same frequency band. At the receiver, the functional value of the aggregated signals is directly calculated using the waveform superposition property of wireless channels, instead of first decoding the individual data stream from each device. There has been comprehensive research for the efficient implementation of AirComp, including the design of beamforming in multi-input-multi-output (MIMO) system (see, e.g., \cite{chen2018over,Zhu2019IoTJ,li2019wireless,wen2019reduced,zhai2021hybrid}), power control for combating the non-uniform channel fading (see, e.g., \cite{cao2020optimized}), the investigation of tradeoff between computation and energy efficiency (see, e.g., \cite{liu2020over}), the design of unmanned aerial vehicle (UAV) assisted AirComp (see, e.g., \cite{fu2021uav}), etc. In view of its low-latency merit in wireless data aggregation, AirComp has been a promising technique widely exploited in federated edge learning for communication efficiency enhancement (see, e.g., \cite{zhu2020broadband,yang2020federated,amiri2020federated,sun2022dynamic,liu2021privacy}). Most recently, researchers have proposed an AirComp based edge inference system, where the same inference task is performed in multiple devices and a server aggregates the local inference results and makes a final decision based on majority voting \cite{yilmaz2022over}. However, such existing design builds on the traditional on-device inference, which causes huge computation overhead at the resource-limited edge devices. It remains an uncharted area to implement edge split inference using AirComp, and thus motivates the current work.

In the considered multi-device edge inference system, the server is equipped with multiple antennas and all devices are equipped with one single antenna. The server aggregates all local feature vectors in a low-latency manner via AirComp to attain a denoised feature vector for the subsequent inference task. In such a system, the transmit precoding and receive beamforming need to be jointly designed to rein in the aggregation error caused by AirComp and maximize the inference accuracy. 
\begin{figure}
\centering
\includegraphics[width=0.5\textwidth]{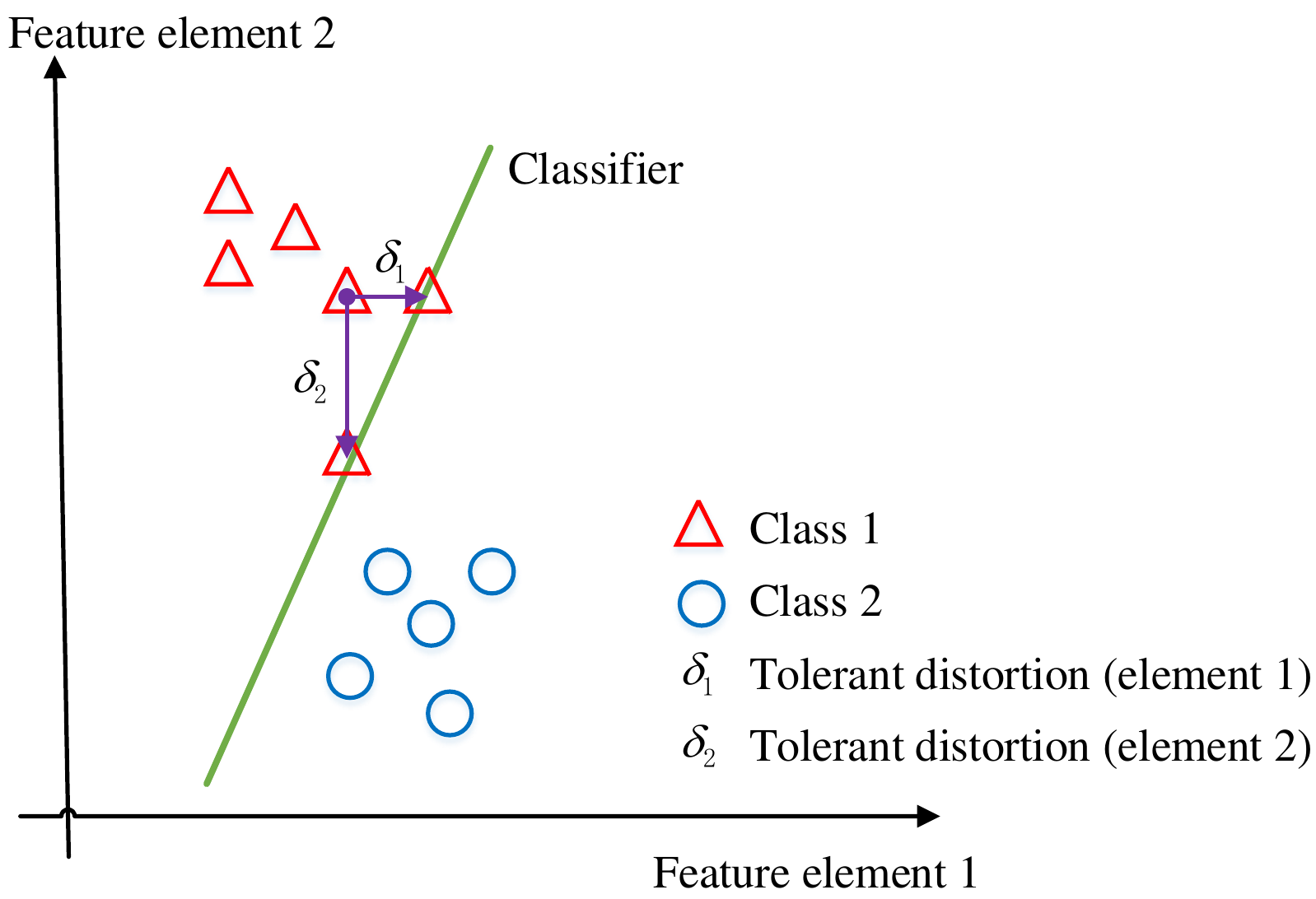}
\caption{Various distortion tolerance of different feature elements in classification tasks: For a distortion level $\delta_1$ obtained under the MMSE criterion, incorrect inference occurs if it is on element 1, but the inference result is correct if it is on element 2.} \label{Fig:DistortionLevel}
\end{figure}
It is noteworthy that the traditional AirComp design criterion, i.e., minimum mean square error (MMSE) used in existing literature, is no longer effective for edge split inference systems. To be specific, the schemes based on MMSE minimize the average distortion between aggregated data by AirComp and the ideally aggregated one without any corruption by channel fading and noise. However, in the context of edge inference, the MSE measure fails to respect the fact that some feature dimensions are more sensitive to the aggregation error than the others when the ultimate inference accuracy is concerned.  As an example, a classification task is shown in Fig. \ref{Fig:DistortionLevel}, whose feature vector has two elements (dimensions). It is observed that feature element 2 is more tolerant to distortion than element 1 in terms of violating the inference accuracy. Obviously, in the case of MMSE, the non-uniform importance levels at different feature elements are ignored and thus may lead to poor performance. To address this issue, the best approach is directly maximizing the inference accuracy in the AirComp design. However, it is not tractable to measure the instantaneous inference accuracy prior to the execution of the inference task. Alternatively, this work adopts an approximate but tractable metric, namely discriminant gain, as the surrogate accuracy measure for classification tasks. The metric is originally proposed in \cite{lan2021progressive} building on the well-known KL divergence \cite{kullback1951information}. Specifically, for arbitrary two classes in the Euclidean feature space, discriminant gain is the distance of their centroids normalized by their covariance. With a larger distance, the two classes are better separated, which implies a higher inference accuracy. However, the joint design of transmit precoding and receive beamforming in AirComp under the criterion of discriminant gain maximization still faces the challenges due to the complicated form of the objective function, and the coupling between the control variables. 

To address the challenges above, the solution framework for inference-task-oriented AirComp is proposed in this paper. The detailed contributions are summarized below.
\begin{itemize}
    \item {\bf Multi-device Over-the-air Inference Systems}: An AirComp based multi-device edge split inference system is established, where the feature vector used for inference at the server is estimated by aggregating all noisy local feature vectors via AirComp. In each time slot, two real feature elements are linearly analog modulated into a complex scalar symbol;  feature vectors of different devices are transmitted simultaneously as blocks of symbols. Under the system settings, the impact of the sensing noise and channel noise on the inference accuracy is theoretically characterized by the derived discriminant gain in closed-form.

    \item {\bf Task-oriented AirComp Transceiver Design for Edge Inference}: Based on the derived discriminant gain, a joint transmit precoding and receive beamforming design problem for the over-the-air inference system is formulated as a problem of maximizing the discriminant gain. To tackle the challenging non-convex problem, the method of  variables transformation is first applied to convert it to an equivalent difference of convex (d.c.) form, which is further solved via using the technique of successive convex approximation (SCA) (see e.g., \cite{razaviyayn2014successive}) to yield a sub-optimal solution. 
    \begin{itemize}
    \item It is noteworthy that the sub-optimal solution meet all the Karush -- Kuhn -- Tucker (KKT) conditions of the original problem, from which one can derive the insight that the optimal beam steered by a device to the edge server has the power inversely proportional to the sensing noise incurred by the device. This further suggests that optimal joint beamforming design should favor those devices with good sensing quality (i.e., small sensing noise).
    \end{itemize}
    
    \item {\bf Performance Evaluation}: Extensive simulations using the wireless sensing simulator proposed in \cite{Li2021SPAWC} have been performed while taking into account the specific task of wide-view human motion recognition with two inference models, i.e., support vector machine (SVM) and multi-layer perception (MLP) neural network, respectively. It is demonstrated that, for both models, maximizing the discriminant gain is effective in maximizing the inference accuracy. Furthermore, it is demonstrated that the proposed scheme significantly outperforms the benchmarking scheme (designed using the criteria of MMSE) in terms of inference accuracy. 
    
\end{itemize}


\section{System Model}
\subsection{Network and Sensing Model}
\begin{figure}
\centering
\includegraphics[width=0.85\textwidth]{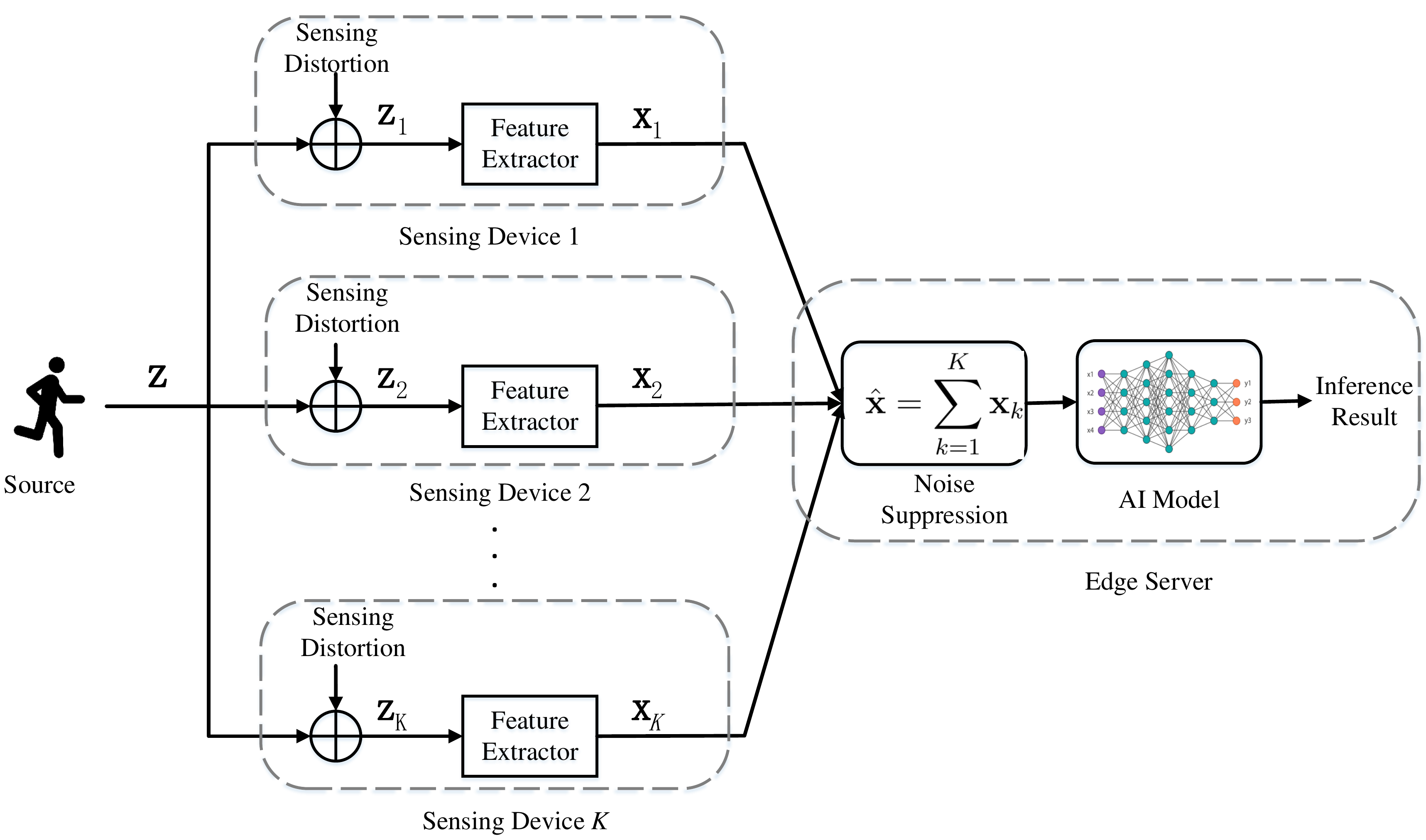}
\caption{Model of Over-the-air Computation Based Edge Inference Systems.}\label{Fig:AirCompApp}
\end{figure}
Consider an edge inference system where there is one server equipped with a multi-antenna access point (AP) and $K$ single-antenna sensing devices (e.g., radar sensors and cameras), as shown in Fig. \ref{Fig:AirCompApp}. The server aims at aggregating the noisy local feature vectors, which are extracted from the real-time noise-corrupted sensory data, on all devices to form a global denoised feature vector for completing the remaining inference task. 
Specifically, the noise-corrupted sensory data obtained by device $k$ is given as
\begin{equation}
{\bf z}_k = {\bf z} + {\bf e}_k,
\end{equation}
where ${\bf z}=[z_1,z_2,...,z_S]^T$ is the ground-true sensory data of the source, ${\bf z}_k = [z_{k,1},z_{k,2},...,z_{k,S}]^T$ is the local observation of device $k$, ${\bf e}_k$ is the sensing noise, and $S$ is the dimension of the raw sensory data. According to  \cite{luo2005universal,xiao2006power}, different elements of the sensing noise vector follow identical and independent zero-mean Gaussian distributions:
\begin{equation}\label{Eq:SensingNoiseDistribution1}
{\bf e}_k \sim \mathcal{N}\left({\bf 0},  \epsilon_k^2 {\bf I}  \right),
\end{equation} 
where $\mathcal{N}(\cdot,\cdot)$ is the Gaussian distribution, $\epsilon_k^2$ is the sensing noise power, and ${\bf I}\in \mathbb{R}^{S\times S} $ is the identical matrix.

The server and the sensing devices communicate via wireless links. Time-division multiple access is adopted. The channels are assumed to be static in each time slot and varying among different slots. 
The channel gain of device $k$ is denoted as ${\bf h}_k \in \mathbb{C}^{N}$, with $N$ being the number of receive antennas at the server and $\mathbb{C}^N$ being a complex vector space with the dimension of $N$. Moreover, the server is assumed to work as the coordinator and has the ability to acquire the channel gains of all devices' uplink links. 


\subsection{Feature Generation and Distribution}
In this part, the feature generation procedure is first introduced, followed by the description of the feature distribution.

\subsubsection{Feature Generation}
 As transmitting the raw sensory data with large dimensions causes large communication overhead as well as violates the data privacy, an alternative approach is to move the feature extraction part (e.g., PCA and convolutional operations) of an AI model on devices. In this work, PCA is adopted to extract a latent low-dimensional feature sub-space from the raw sensory data on  each device. The detailed procedure is described as follows.
\begin{itemize}
\item At the training stage, PCA is first performed by the server over the offline training dataset to extract the principal dimensions of each sample.  The learning model is trained  using the principal feature dimensions.
\item At the inference stage, before the server aggregates the local observations from each device, the principal eigen-space is broadcast to each device. For each device, the local feature vector is extracted by projecting the sensory data into the principal eigen-space, and then transmitted. 
\end{itemize} 
Thereby, the extracted local feature vectors from the sensory data ${\bf z}_k$ can be expressed as 
\begin{equation}\label{Eq:Observation}
\begin{aligned}
{\bf x}_k = {\bf U}^T {\bf z}_k & = {\bf U}^T{\bf z} + {\bf U}^T{\bf e}_k  ={\bf x} + {\bf d}_k, \quad 1\leq k \leq K,
\end{aligned}
\end{equation}
where ${\bf U}$ is a $S\times M$ real column unitary matrix representing the principal eigen-space of PCA, $M$ is the dimension of the principal feature eigen-space, 
\begin{equation}
    {\bf x}={\bf U}^T{\bf z} = [x_1,x_2,...,x_M]^T,
\end{equation} is the ground-true feature vector, and  
\begin{equation}
    {\bf d}_k = {\bf U}^T{\bf e}_k,\quad 1\leq k \leq K,
\end{equation} 
is the projected noise vector of device $k$. By substituting the distribution of ${\bf e}_k$ in \eqref{Eq:SensingNoiseDistribution1}, the distribution of ${\bf d}_k$ can be derived as
\begin{equation}\label{Eq:SensingNoiseDistribution}
    {\bf d}_k \sim \mathcal{N}\left({\bf 0}, \epsilon_k^2 {\bf I} \right), \quad 1\leq k \leq K,
\end{equation}
where the variance is derived from
$    \mathbb{E}\left({ \bf d}_k^T {\bf d}_k \right) = \mathbb{E}\left( {\bf U}^T {\bf e}_k{\bf e}_k^T {\bf U} \right) = {\bf U}^T \mathbb{E}( {\bf e}_k {\bf e}_k^T) {\bf U} = \epsilon_k^2 {\bf I}$.

\subsubsection{Feature Distribution}
By considering a classification inference task with $L$ classes and following the setting in \cite{lan2021progressive}, the ground-true feature vector ${\bf x}$ is assumed to follow a Gaussian mixture as
\begin{equation}\label{Eq:GroundTrueDistribution}
\mathcal{F}({\bf x}) = \dfrac{1}{L}\sum\limits_{\ell=1}^L \mathcal{F}_{\ell} ({\bf x}),
\end{equation}
where $\mathcal{F}_{\ell} ({\bf x})$ is the Gaussian distribution of ${\bf x}$ in terms of the $\ell$-th class. As PCA is performed, different feature elements are linearly independent. Thereby, $\mathcal{F}_{\ell} ({\bf x})$ can be written as
\begin{equation}\label{Eq:GroundTrueClassDistribution}
     \mathcal{F}_{\ell} ({\bf x}) = \mathcal{N}\left({\bm \mu}_{\ell}, {\bm \Sigma}\right), \quad 1 \leq \ell \leq L,
\end{equation}
where  ${\bm \mu}_{\ell} \in \mathbb{R}^M$ is the centroid of the $\ell$-th class, given as
\begin{equation}\label{Eq:Centroid}
 {\bm \mu}_{\ell} = [\mu_{\ell,1},\mu_{\ell,2},...,\mu_{\ell,M}]^T, \quad 1\leq \ell \leq L,
\end{equation}
and $ {\bm \Sigma} \in \mathbb{R}^{M\times M}$ is a diagonal covariance matrix, given as
\begin{equation}\label{Eq:Covariance}
 {\bm \Sigma} = \diag\{\sigma_1^2,\sigma_2^2,...,\sigma_M^2\}.
\end{equation}

\subsection{Inference Capability}
\begin{figure}
\centering
\includegraphics[width=0.45\textwidth]{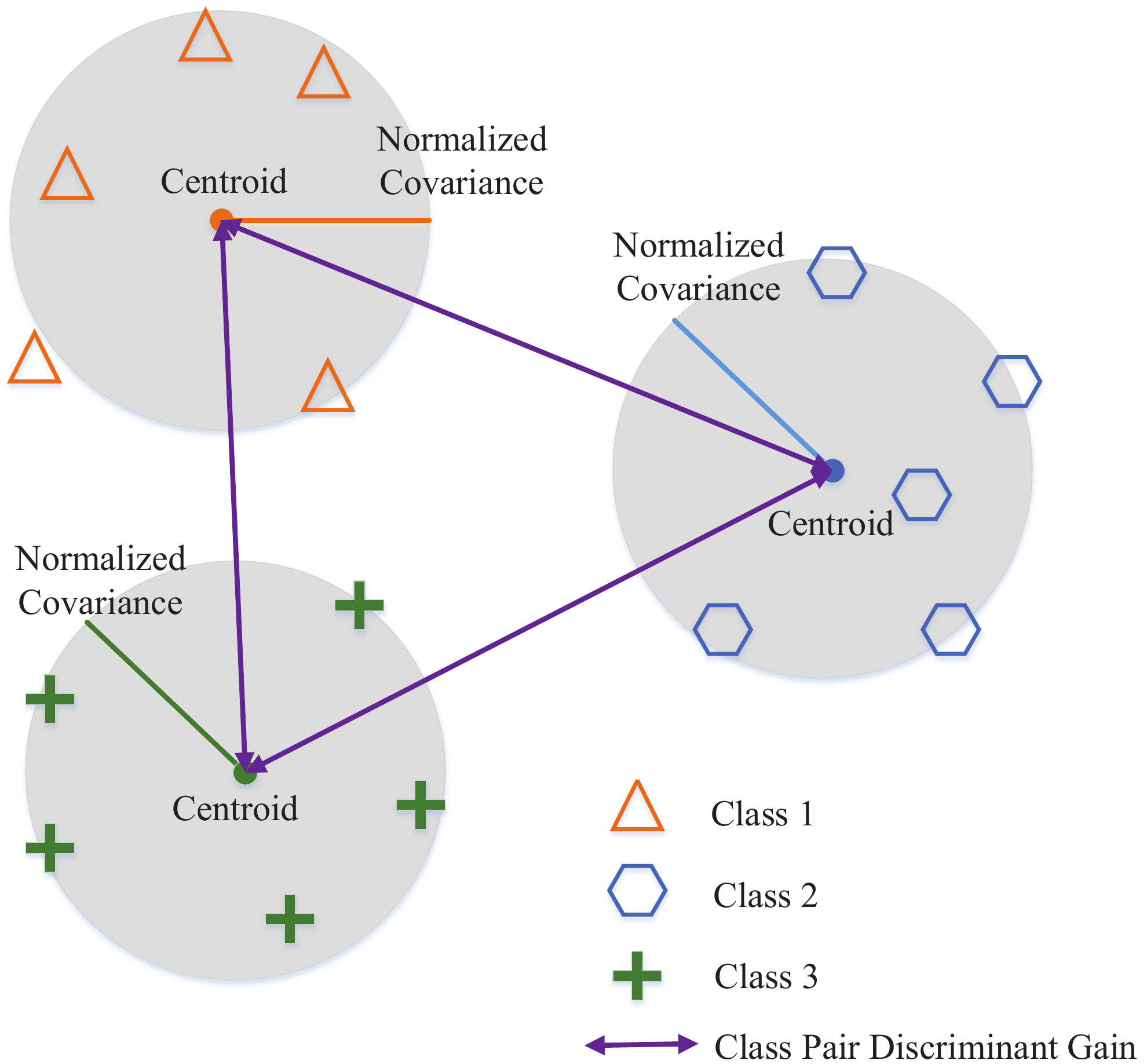}
\caption{Geometric interpretation of discriminant gain in the feature space.}\label{Fig:DG-FeatureSpace}
\vspace{-1em}
\end{figure}

In this work, the metric \emph{discriminant gain} proposed in \cite{lan2021progressive} is adopted as the inference accuracy measure for classification tasks.  For arbitrary two classes, the discriminant gain represents the distance  between their centroids in the Euclidean feature space under normalized covariance, as presented in Fig. \ref{Fig:DG-FeatureSpace}. That says, a larger discriminant gain between two classes means that they are more likely to be differentiated, and thus implies a higher inference accuracy. In the sequel, the mathematical model of discriminant gain is introduced.

Discriminant gain is derived from the well-known KL divergence proposed in \cite{kullback1951information}. Consider an arbitrary class pair, say classes $\ell$ and $\ell^{'}$,  and the feature space expanded by the feature vector ${\bf x}$. Based on the distribution of ${\bf x}$ in \eqref{Eq:GroundTrueDistribution} and according to \cite{lan2021progressive}, the pair-wise discriminant gain is defined as 
\begin{equation}\label{Eq:PairWiseGain}
\begin{aligned}
G_{\ell,\ell^{'}}({\bf x}) &= \mathsf{D}_{KL}\left[\mathcal{F}_{\ell}({\bf x})\; \big\| \; \mathcal{F}_{\ell^{'}}({\bf x}) \right] + \mathsf{D}_{KL}\left[\mathcal{F}_{\ell^{'}}({\bf x})\; \big\| \; \mathcal{F}_{\ell}({\bf x}) \right],\\
& = \int_{\bf x} \mathcal{F}_{\ell}({\bf x}) \log\left[ \dfrac{\mathcal{F}_{\ell^{'}}({\bf x}) }{\mathcal{F}_{\ell}({\bf x})} \right] {\rm d}{\bf x} +\int_{\bf x} \mathcal{F}_{\ell^{'}}({\bf x}) \log\left[ \dfrac{\mathcal{F}_{\ell}({\bf x}) }{\mathcal{F}_{\ell^{'}}({\bf x})} \right] {\rm d}{\bf x},\\
& = \left( {\bm \mu}_{\ell} -{\bm \mu}_{\ell^{'}}\right)^T {\bm \Sigma}^{-1} \left(  {\bm \mu}_{\ell} -{\bm \mu}_{\ell^{'}} \right),\quad \forall (\ell,\ell^{'}),
\end{aligned}
\end{equation}
where $\mathsf{D}_{KL}(\cdot \| \cdot)$ is the KL divergence defined in \cite{kullback1951information}. As different feature elements are independent, it follows that
\begin{equation}
    G_{\ell,\ell^{'}}({\bf x}) = \sum\limits_{m=1}^M G_{\ell,\ell^{'}}(x_m),
\end{equation}
where $x_m$ is the $m$-th element in ${\bf x}$ and $G_{\ell,\ell^{'}}(x_m)$ is given as
\begin{equation}\label{Eq:PairWiseGainDiemsnion}
G_{\ell,\ell^{'}}(x_m) = \dfrac{\left( \mu_{\ell,m} - \mu_{\ell^{'},m}\right)^2}{\sigma_m^2}, \quad 1\leq m \leq M,
\end{equation}
and the other notations follow that in \eqref{Eq:Centroid} and \eqref{Eq:Covariance}. Then, the overall discriminant gain is defined as the average of all pair-wise discriminant gains in \eqref{Eq:PairWiseGain}, given as
\begin{equation}\label{Eq:DiscriminantGain}
\begin{aligned}
G({\bf x})  & = \dfrac{2}{L(L-1)} \sum\limits_{\ell^{'} =1}^L \sum\limits_{\ell<\ell^{'} } G_{\ell,\ell^{'}} ({\bf x}) =   \dfrac{2}{L(L-1)} \sum\limits_{\ell^{'} =1}^L \sum\limits_{\ell<\ell^{'} } \sum\limits_{m=1}^MG_{\ell,\ell^{'}}(x_m) = \sum\limits_{m=1}^M G( x_m), 
\end{aligned}
\end{equation}
where $G( x_m)$ is the discriminant gain of the $m$-th feature elements, given as
\begin{equation}\label{Eq:OneDimensionDiscriminantGain}
G( x_m) = \dfrac{2}{L(L-1)}  \sum\limits_{\ell^{'} =1}^L \sum\limits_{\ell<\ell^{'} } \dfrac{\left( \mu_{\ell,m} - \mu_{\ell^{'},m}\right)^2}{\sigma_m^2}, \quad 1\leq m \leq M.
\end{equation}

\subsection{AirComp Model}
The technique of AirComp is used to aggregate the local feature vectors $\{{\bf x}_k\}$ from all devices, as it can suppress the sensing noise and significantly enhance the communication efficiency. Specifically, each device transmits a complex scalar symbol via the single antenna in each time slot. 
The real part and the imaginary part of the complex scalar symbol contain one feature element, respectively. At the server, AirComp is performed to aggregate the two feature elements and estimate their ground-true versions. Thereby, the whole feature vector can be grouped into different element pairs, which can be sequentially transmitted in a time-division way over several time slots. Obviously, the design of AirComp in all time slots is the same. Without loss of generality, in the sequel, the transmission in an arbitrary time slot is considered.

\begin{figure}
\centering
\includegraphics[width=0.9\textwidth]{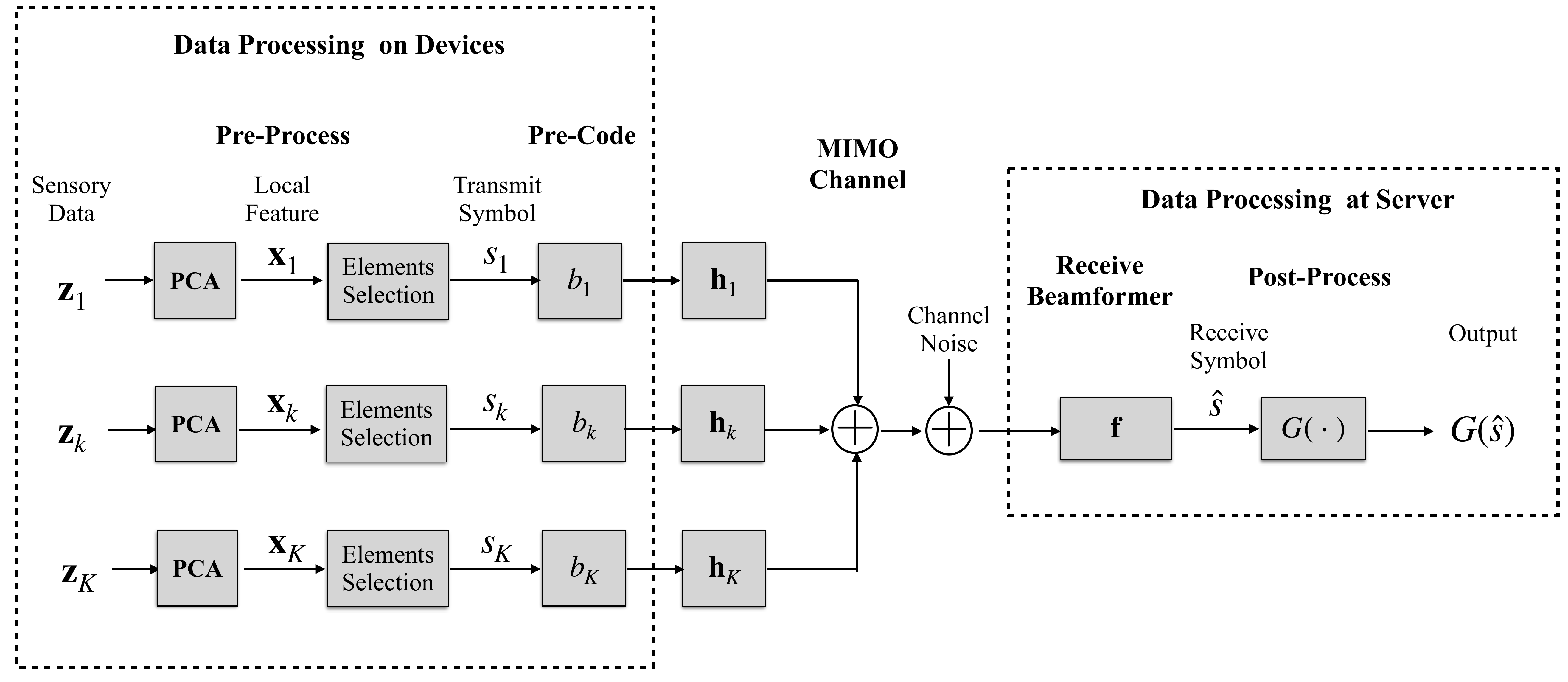}
\caption{Block diagram of AirComp for Feature Aggregation.}\label{Fig:AirCompModel}
\vspace{-1em}
\end{figure}
Consider the case where the server aggregates the $m_1$-th and $m_2$-th local elements from all devices,  say $\{x_{k,m_1}, x_{k,m_2},\;1\leq k\leq K\}$, to estimate the ground-true feature elements $\{x_{m_1},x_{m_2}\}$. The procedure of AirComp is shown in Fig. \ref{Fig:AirCompModel} and is described as follows. For an arbitrary device $k$, its local sensory data is first pre-processed by PCA to extract the principal feature elements. Then, the $m_1$-th and $m_2$-th principal feature dimensions, say $x_{k,m_1}$ and $x_{k,m_2}$, are combined in one symbol for transmission, as
\begin{equation}\label{Eq:TransmitSymbol}
s_k = x_{k,m_1} + j x_{k,m_2},\quad 1\leq k \leq K,
\end{equation}
where $s_k \in \mathbb{C}$ is the transmitted symbol and $j$ represents the imaginary unit. Next, $s_k$ is further pre-coded with a scalar $b_k \in \mathbb{C}$ and transmitted over a MIMO channel. At the server, the receive signal is the aggregation of all transmit symbols, given as
\begin{equation}\label{Eq:ReceiveSignal}
{\bf y}_m = \sum\limits_{k=1}^K {\bf h}_k b_k s_k+ {\bf n},
\end{equation}
where 
${\bf n}$ is the additive white Gaussian noise with the following distribution:
\begin{equation}\label{Eq:NoiseDistribution}
{\bf n} \sim \mathcal{N}\left( {\bf 0}, \delta_0^2 {\bf I} \right),
\end{equation}
and $\delta_0^2$ is the noise variance. Next, a receive beamforming vector ${\bf f} \in \mathbb{C}^N$ is used to aggregate all local symbols $\{s_k\}$ to generate the estimates of the ground-true feature elements $x_{m_1}$ and $x_{m_2}$. Specifically, the received symbol after receive beamforming can be written as
\begin{equation}\label{Eq:EstimateSymbol}
\hat{s} =  {\bf f}^H {\bf y}_m = {\bf f}^H\sum\limits_{k=1}^K {\bf h}_k b_k s_k + {\bf f}^H{\bf n}.
\end{equation}
It follows that the estimates are given by
\begin{equation}\label{Eq:Estimate}
\left\{
\begin{aligned}
& \hat{x}_{m_1} = \mathsf{Re}\left( \hat{s} \right)  = \mathsf{Re}\left( {\bf f}^H\sum\limits_{k=1}^K {\bf h}_k b_k s_k + {\bf f}^H{\bf n}\right),\\
&  \hat{x}_{m_2} = \mathsf{Im}\left( \hat{s} \right)  = \mathsf{Im}\left( {\bf f}^H\sum\limits_{k=1}^K {\bf h}_k b_k s_k + {\bf f}^H{\bf n}\right),
\end{aligned}
\right.
\end{equation}
where  $\hat{x}_{m_1}$ and $\hat{x}_{m_2}$ are the estimates of  $x_{m_1}$ and $x_{m_2}$ respectively,  $ \mathsf{Re}(\cdot)$ and $ \mathsf{Im}(\cdot)$ are the functions to extract real part and imaginary part of one complex number respectively, and other notations follow that in \eqref{Eq:ReceiveSignal}. Finally, $\hat{x}_{m_1}$ and $\hat{x}_{m_2}$ are post-processed to output the discriminant gain $G(\hat{x}_{m_1}) + G(\hat{x}_{m_2}) $. 

\section{Problem Formulation and Simplification}
\subsection{Problem Formulation}
Different from the traditional AirComp design, which aims at minimizing the distortion between the estimated feature elements $\{ \hat{x}_{m_1},\hat{x}_{m_2} \}$ and the ground-true ones $\{x_{m_1},x_{m_2}\}$ without taking into account the performance metric of the specific tasks, in this work, the design objective follows the task-oriented principle and maximizes the inference accuracy measured by the sum discriminant gains of $\hat{x}_{m_1}$ and $\hat{x}_{m_2}$, given as 
\begin{equation}\label{Eq:Objective}
\max\; G = G(\hat{x}_{m_1}) + G(\hat{x}_{m_2}),
\end{equation}
where $\hat{x}_{m_1}$  and $\hat{x}_{m_2}$ defined in \eqref{Eq:Estimate} are the estimates of the ground-true feature elements, and $G(\hat{x}_{m_1})$ and $G(\hat{x}_{m_2})$ are the corresponding discriminant gains. 

Besides, there is one constraint on the transmit power of each device, given by
\begin{equation}\label{Eq:Constraint}
b_k\mathbb{E}(s_ks_k^H)b_k^H\leq P_k, \quad 1\leq k \leq K,
\end{equation}
where  $b_k$ is the precoding scalar at device $k$, $b_k^H$ is the hermitian of $b_k$, $s_k$ is the transmit symbol, and $P_k$ is the total transmit power of device $k$. 
The transmit symbol variance, say $\mathbb{E}(s_ks_k^H)$, can be estimated from the offline training data samples, and thus is known by the edge server as prior information. Therefore, the power constraint in \eqref{Eq:Constraint} can be re-written as
\begin{equation}\label{Eq:Constraint1}
b_k b_k^H \leq \hat{P}_k,\quad 1\leq k \leq K,
\end{equation}
where $\hat{P}_k$ is the maximum transmit precoding power, given as
\begin{equation}\label{Eq:TransmitPrecodingPower}
  \hat{P}_k = \dfrac{ P_k }{ \mathbb{E}(s_ks_k^H) },\quad 1\leq k \leq K.
 \end{equation}


In summary, the discriminant gain maximization problem can be written as
\begin{equation}\text{(P1)}\qquad
\begin{aligned}
\max\limits_{\{b_k\},{\bf f}}&\;\; G = G(\hat{x}_{m_1}) + G(\hat{x}_{m_2}),\\
\text{s.t.} &\;\; b_k b_k^H\leq \hat{P}_k, \quad 1\leq k \leq K.
\end{aligned}
\end{equation}
The formulation of (P1) follows the task-oriented principle. To be specific, the inference accuracy measured by the discriminant gain is maximized instead of using the MMSE criterion. That's because the MMSE criterion ignores the fact that a same distortion level on different feature elements has different impacts on the inference accuracy, and thus leads to poor performance. The task-oriented formulation, however, causes new challenges. To begin with, the discriminant gain has a complicated non-convex sum-of-ratios form. Besides,  the design of the receive beamforming ${\bf f}$ and the precoding scalars $\{b_k\}$ are coupled [see \eqref{Eq:Estimate}]. Moreover, the feature elements in the received symbol are cross coupled, i.e., each estimated feature element defined in \eqref{Eq:Estimate} could be a linear combination of the  ground-true elements $x_{m_1}$ and $x_{m_2}$, due to channel rotation.  This leads to a complicated distribution of $\hat{x}_{m_1}$ and $\hat{x}_{m_2}$, and thus a complicated expression of the discriminant gains $G(\hat{x}_{m_1})$ and $G(\hat{x}_{m_2})$. 

\subsection{Discriminant Gains with Zero-Forcing Pre-coders}
To address the challenges mentioned above, we simplify (P1) with two steps in this part. The well-known zero-forcing (ZF) precoders are first used to simplify the estimated feature elements $\{\hat{x}_{m_1}, \hat{x}_{m_2}\}$. Then, based on the ZF precoders, the  discriminant gains, i.e., $G(\hat{x}_{m_1})$ and $G(\hat{x}_{m_2})$, are derived to simplify the objective function.

\subsubsection{ZF precoders}
First, the ZF design is given by
\begin{equation}\label{Eq:ZF}
{\bf f}^H {\bf h}_k b_k = c_k, \quad 1\leq k \leq K,
\end{equation}
where ${\bf f}^H$ is the receive beamforming vector, $ {\bf h}_k$ is the channel vector of device $k$, $ b_k$ is the precoder of device $k$, and  $c_k \geq 0$ is a real number representing the receive signal strength, or called steering power, from device $k$. Then, the ZF precoders can be derived as
\begin{equation}\label{Eq:Precoder}
b_k = \dfrac{ c_k {\bf h}_k^H{\bf f} }{  {\bf h}_k^H {\bf f} {\bf f}^H {\bf h}_k }, \quad 1 \leq k \leq K.
\end{equation}
It follows that the power constraint in (P1) can be re-written as
\begin{equation}\label{Eq:ReceiveStrengthConstraint}
c_k^2 \leq \hat{P}_k  {\bf h}_k^H {\bf f} {\bf f}^H {\bf h}_k, \quad 1\leq k \leq K.
\end{equation}

Besides, by substituting the precoders in \eqref{Eq:Precoder} and $\hat{s}_k$ in \eqref{Eq:TransmitSymbol} into the estimates $\hat{x}_{m_1}$ and $\hat{x}_{m_2}$ in  \eqref{Eq:Estimate}, we can obtain
\begin{equation}\label{Eq:SimplifiedEstimate}
\left\{
\begin{aligned}
& \hat{x}_{m_1} = \mathsf{Re}\left( \sum\limits_{k=1}^K c_k s_k + {\bf f}^H {\bf n}\right) =\sum\limits_{k=1}^K c_k x_{k,m_1} + \mathsf{Re}({\bf f}^H {\bf n}) ,\\
& \hat{x}_{m_2} =\mathsf{Im}\left( \sum\limits_{k=1}^K c_k s_k + {\bf f}^H {\bf n}\right) =\sum\limits_{k=1}^K c_k x_{k,m_2} + \mathsf{Im}({\bf f}^H {\bf n}),
\end{aligned} 
\right.
\end{equation}
where the notations follow that in \eqref{Eq:TransmitSymbol}, \eqref{Eq:Estimate}, and \eqref{Eq:Precoder}.

\subsubsection{Discriminant Gains}
To achieve the discriminant gain $G$, in the sequel, the distributions of the local transmit feature elements $\{x_{k,m_1}, x_{k,m_2}\}$ are first derived. Then, based on the ZF precoders, the distribution of the received elements  $\{\hat{x}_{m_1}, \hat{x}_{m_2}\}$ are derived. Next, the discriminant gains are obtained, followed by the derivation of a simplified problem of (P1).

First, recall the local elements $x_{k,m_1}$ and $x_{k,m_2}$ are given by 
\begin{equation}\label{Eq:ObeservationElement}
x_{k,m_i} = x_{m_i} + d_{k,m_i}, \quad i=1,2, \; 1\leq k \leq K,
\end{equation} 
where the distribution of the ground-true element $x_{m,i}$ is given by 
\begin{equation}\label{Eq:xmi}
x_{m_i} \sim \dfrac{1}{L}\sum\limits_{\ell =1 }^L \mathcal{N}\left(\mu_{\ell,m_i}, \sigma_{m_i}^2 \right), \quad i=1,2,
\end{equation}
according the distribution of ${\bf x}$ in \eqref{Eq:GroundTrueDistribution}, \eqref{Eq:Centroid}, and \eqref{Eq:Covariance}, and the distribution of the noise $d_{k,m_i}$ is given by 
\begin{equation}\label{Eq:dkmi}
d_{k,m_i} \sim \mathcal{N}\left(0,\epsilon_k^2 \right), \quad i=1,2,
\end{equation}
according to the distribution of ${\bf d}_k$ in \eqref{Eq:SensingNoiseDistribution}. Subsequently, the following lemma in terms of $x_{k,m_i}$'s distribution can be obtained.
\begin{lemma}\label{Lma:LocalDistribution}
The distribution of the local elements $\{x_{k,m_i}\}$ can be derived as
\begin{equation}\label{Eq:LocalDistribution}
x_{k,m_i} \sim \dfrac{1}{L}\sum\limits_{\ell =1 }^L \mathcal{N}\left(\mu_{\ell,m_i}, \sigma_{m_i}^2 + \epsilon_k^2 \right), \quad i=1,2,\; 1\leq k \leq K,
\end{equation}
\end{lemma}
\proof Please see Appendix \ref{Apdx:LmaLocalDistribution}.

Then, by substituting the distributions of $\{x_{k,m_1},x_{k,m_2}\}$ in \eqref{Eq:LocalDistribution} and the distribution of the channel noise ${\bf n}$ in \eqref{Eq:NoiseDistribution} into the received feature elements $\{\hat{x}_{m_1},\hat{x}_{m_2}\}$ in \eqref{Eq:SimplifiedEstimate}, their distributions can be derived as shown in Lemma \ref{Lma:EstimateDistribution}.
\begin{lemma}\label{Lma:EstimateDistribution}
The distribution of the estimated feature elements $\{\hat{x}_{k,m_i}\}$ are given by
\begin{equation}\label{Eq:EstimateDistribution}
\hat{x}_{m_i} \sim \dfrac{1}{L}\mathcal{N}\left( \hat{\mu}_{\ell,m_i}, \hat{\sigma}_{m_i}^2\right), \quad i = 1,2,
\end{equation}
where the centroids $ \{ \hat{\mu}_{\ell,m_i} \} $ and the variance $\{ \hat{\sigma}_{m_i}^2 \}$ are
\begin{equation}\label{Eq:EstimateMeanCovariance}
\left\{
\begin{aligned}
& \hat{\mu}_{\ell,m_i} = \sum\limits_{k=1}^K c_k \mu_{\ell,m_i},\;\; i = 1,2,\\
&\hat{\sigma}_{m_i}^2 = \sigma_{m_i}^2 \left(  \sum\limits_{k=1}^K c_k \right)^2+ \sum\limits_{k=1}^K c_k^2 \epsilon_k^2 + \dfrac{\delta_0^2}{2}  \left( {\bf f}_1^T {\bf f}_1 +  {\bf f}_2 {\bf f}_2 \right),\;\; i = 1,2,\\
\end{aligned}
\right.
\end{equation}
$\delta_0^2$ is the channel noise power, ${\bf f}_1 = \mathsf{Re}({\bf f})$ and ${\bf f}_2 = \mathsf{Im}({\bf f})$ are the real part and imaginary part of the receive beamforming ${\bf f}$ respectively, and  other notations follow that in \eqref{Eq:xmi} and \eqref{Eq:dkmi}.
\end{lemma}
\proof Please see Appendix \ref{Apdx:LmaEstimateDistribution}.

Next, based on the distributions in Lemma \ref{Lma:EstimateDistribution} and the definition of discriminant gain in \eqref{Eq:OneDimensionDiscriminantGain}, the discriminant gains of $\{x_{k,m_1},x_{k,m_2}\}$ can be derived as
\begin{equation}\label{Eq:Gxmi}
G(\hat{x}_{m_i}) =  \dfrac{2}{L(L-1)}  \sum\limits_{\ell^{'} =1}^L \sum\limits_{\ell<\ell^{'} } \dfrac{\left( \hat{\mu}_{\ell,m_i} - \hat{\mu}_{\ell^{'},m_i}\right)^2}{\hat{\sigma}_{m_i}^2},\quad i=1,2,
\end{equation}
where $\{\hat{\mu}_{\ell,m_i}\}$  and $\{\hat{\sigma}_{m_i}^2\}$ are defined in \eqref{Eq:EstimateMeanCovariance}.

Finally, by substituting the discriminant gains of $\{x_{k,m_1},x_{k,m_2}\}$ in \eqref{Eq:Gxmi} and the power constraint in \eqref{Eq:ReceiveStrengthConstraint} into  (P1), it can be equivalently derived as
\begin{equation}\text{(P2)}\qquad
\begin{aligned}
\max\limits_{\{c_k\},{\bf f}_1 ,{\bf f}_2 }&\;\; G =  \dfrac{2}{L(L-1)} \sum\limits_{i=1}^2 \sum\limits_{\ell^{'} =1}^L \sum\limits_{\ell<\ell^{'} } \dfrac{\left( \hat{\mu}_{\ell,m_i} - \hat{\mu}_{\ell^{'},m_i}\right)^2}{\hat{\sigma}_{m_i}^2},\\
\text{s.t.} &\;\; c_k^2 \leq \hat{P}_k  {\bf h}_k^H \left( {\bf f}_1{\bf f}_1^T + {\bf f}_2{\bf f}_2^T \right) {\bf h}_k, \quad 1\leq k \leq K,
\end{aligned}
\end{equation}
where the notations follow that in \eqref{Eq:EstimateMeanCovariance}.

\section{Joint Power Control and Receive Beamforming for Task-oriented AirComp}
In this section, variables transformation is first applied to derive (P2) into an equivalent d.c. problem. Then, the method of SCA is adopted to address it and obtain the joint design of steering power control and receive beamforming. Finally, the struction of the obtained solution is investigated.

\subsection{An Equivalent D.C. Problem}
In this part, to simplify (P2), the following variables are first defined:
\begin{equation}\label{Eq:DG_Noise}
    \alpha_{\ell,\ell^{'},m_i} = \dfrac{\left( \hat{\mu}_{\ell,m_i} - \hat{\mu}_{\ell^{'},m_i}\right)^2}{\hat{\sigma}_{m_i}^2},\quad \forall (\ell,\ell^{'},m_i),
\end{equation}
where $\alpha_{\ell,\ell^{'},m_i}$ represents the per class pair discriminant gain of the $m_i$-th received element $\hat{x}_{m_i}$. 
It follows that (P2) can be equivalently derived as
\begin{equation}\label{Eq:MP1}
\begin{aligned}
\max\limits_{\{c_k\},{\bf f}_1 ,{\bf  f}_2,\left\{\alpha_{\ell,\ell^{'},m_i} \right\}  }\;\;& G =  \dfrac{2}{L(L-1)} \sum\limits_{i=1}^2 \sum\limits_{\ell^{'} =1}^L \sum\limits_{\ell<\ell^{'} } \alpha_{\ell,\ell^{'},m_i},\\
\text{s.t.} \;\;& c_k^2 \leq \hat{P}_k  {\bf h}_k^H \left( {\bf f}_1{\bf f}_1^T + {\bf f}_2{\bf f}_2^T \right) {\bf h}_k, \quad 1\leq k \leq K,\\
&\left(\hat{\mu}_{\ell,m_i} - \hat{\mu}_{\ell^{'},m_i}\right)^2 = \alpha_{\ell,\ell^{'},m_i} \hat{\sigma}_{m_i}^2 ,\quad \forall (\ell,\ell^{'},m_i),
\end{aligned}
\end{equation}
where 
\begin{equation}\label{Eq:MuDifference}
    \left(\hat{\mu}_{\ell,m_i} - \hat{\mu}_{\ell^{'},m_i}\right)^2 =  \left( \sum\limits_{k=1}^K c_k\right)^2 \left( \mu_{\ell,m_i} - \mu_{\ell^{'},m_i}\right)^2 , \forall (\ell,\ell^{'},m_i),
\end{equation}
and
\begin{equation}\label{Eq:Sigma}
    \hat{\sigma}_{m_i}^2 =  \left[  \sigma_{m_i}^2 \left(  \sum\limits_{k=1}^K c_k \right)^2+ \sum\limits_{k=1}^K c_k^2 \epsilon_k^2 + \dfrac{\delta_0^2}{2} \left( {\bf f}_1^T {\bf f}_1 +  {\bf f}_2^T {\bf f}_2 \right) \right],\; \forall (\ell,\ell^{'},m_i).
\end{equation}

Then, it can be shown that using symmetric real and imaginary receive beamformers can achieve the optimal solution of the probelm in \eqref{Eq:MP1}, as presented in the following lemma.
\begin{lemma}[Symmetric Receive Beamformers]\label{Lma:SymmteicBeamformers}
Symmetric real and imaginary receive beamformers, as  in \eqref{Eq:SymmetricBeamformers}, will not influence the optimality of the problem in \eqref{Eq:MP1}.
\begin{equation}\label{Eq:SymmetricBeamformers}
{\bf f}_1 = {\bf f}_2 = \hat{\bf f}.
\end{equation}
\end{lemma}
\proof Please see Appendix \ref{Apdx:LmaSymmteicBeamformers}.

Besides, it can be further proved that extending the feasible region of the second constraint of the problem in \eqref{Eq:MP1}, i.e., the equality constraint, has no influence on its optimal solution, as equality should be achieved to obtain the optimum, as presented in Lemma \ref{Lma:ExtendedRegion}.
\begin{lemma}[Equivalent Extended Feasible Region]\label{Lma:ExtendedRegion}
A problem, which extends the feasible region of the second constraint of the problem in \eqref{Eq:MP1} as
\begin{equation}\label{Eq:SecondConstraintEqv}
\left( \sum\limits_{k=1}^K c_k \right)^2  \left[\dfrac{ \left( \mu_{\ell,m_i} - \mu_{\ell^{'},m_i}\right)^2 }{\alpha_{\ell,\ell^{'},m_i} } - \sigma_{m_i}^2 \right] \geq   \sum\limits_{k=1}^K c_k^2 \epsilon_k^2 +  \delta_0^2 \hat{\bf f}^T \hat{\bf f} , \quad \forall (\ell,\ell^{'},m_i),
\end{equation}
and keeps the other constraints and the objective function, achieves the same optimal solution to the  problem in \eqref{Eq:MP1}.
\end{lemma}
\proof Please see Appendix \ref{Apdx:LmaExtendedRegion}.

Next, based on Lemmas \ref{Lma:SymmteicBeamformers} and \ref{Lma:ExtendedRegion}, the problem in \eqref{Eq:MP1} can be equally derived as
\begin{equation*}(\text{P3})\;
\begin{aligned}
\max\limits_{\{c_k\},\hat{\bf f} ,\{ \alpha_{\ell,\ell^{'},m_i} \}  }&\;\; G =  \dfrac{2}{L(L-1)} \sum\limits_{i=1}^2 \sum\limits_{\ell^{'} =1}^L \sum\limits_{\ell<\ell^{'} } \alpha_{\ell,\ell^{'},m_i},\\
\text{s.t.} \;\;& c_k^2 -R_k(\hat{\bf f}) \leq 0, \quad 1\leq k \leq K,\\
&   \sum\limits_{k=1}^K c_k^2 \epsilon_k^2 +  \delta_0^2 \hat{\bf f}^T \hat{\bf f} + \sigma_{m_i}^2\left( \sum\limits_{k=1}^K c_k \right)^2 - Q_{\ell,\ell^{'},m_i}\left( \{c_k \},\alpha_{\ell,\ell^{'},m_i} \right)\leq 0 ,\; \forall (\ell,\ell^{'},m_i),
\end{aligned}
\end{equation*}
where $R_k(\hat{\bf f}) $ and $Q_{\ell,\ell^{'},m_i}\left( \{c_k \},\alpha_{\ell,\ell^{'},m_i} \right)$ are the functions defined as
\begin{equation}\label{Eq:RQRelaxation}
\left\{
\begin{aligned}
& R_k(\hat{\bf f}) = 2\hat{P}_k{\bf h}_k^H\hat{\bf f}\hat{\bf f}^T{\bf h}_k, \quad q\leq k \leq K,\\
& Q_{\ell,\ell^{'},m_i}\left( \{c_k \},\alpha_{\ell,\ell^{'},m_i} \right) = \left( \sum\limits_{k=1}^K c_k \right)^2  \times \dfrac{ \left( \mu_{\ell,m_i} - \mu_{\ell^{'},m_i}\right)^2 }{\alpha_{\ell,\ell^{'},m_i} }, \quad \forall (\ell,\ell^{'},m_i).
\end{aligned}
\right.
\end{equation}
Although (P3) is non-convex due to the two constraints therein, it is a d.c. problem as presented in the following lemma.
\begin{lemma}\label{Lma:DCProblem}
(P3) is a d.c. problem, since $ c_k^2$, $R_k(\hat{\bf f}) $, $ \sum\limits_{k=1}^K c_k^2 \epsilon_k^2 +  \delta_0^2 \hat{\bf f}^T \hat{\bf f}+\sigma_{m_i}^2\left( \sum\limits_{k=1}^K c_k \right)^2$, $Q_{\ell,\ell^{'},m_i}\left( \{c_k \},\alpha_{\ell,\ell^{'},m_i} \right)$, and the objective function are differentiable and convex.
\end{lemma} 
\proof See Appendix \ref{Apdx:LmaDCProblem}.

In the sequel, the SCA method is used to get a sub-oprimal solution based on Lemma \ref{Lma:DCProblem}.


\subsection{SCA Based Solution Approach}
In this part, the SCA approach is used to address (P3) for obtaining a sub-optimal solution based on Lemma \ref{Lma:DCProblem} by iterating over the following two steps:
\begin{itemize}
\item \emph{Convex relaxation}: Based on a feasible reference point, (P3) is relaxed into a convex problem, whose feasible region is a subset of that of (P3). Hence, the solution to the relaxed problem is guaranteed to be feasible for (P3).
\item \emph{Reference point updating}: The solution of the relaxed convex problem is used as the new reference point for the next iteration.
\end{itemize} 
This process iterates till convergence and the final result can be guaranteed to satisfy the KKT conditions of (P3) \cite{razaviyayn2014successive}. In the sequel, the approach of convex relaxation is first presented, followed by the summary of the overall joint steering power control and receive beamforming algorithm.

\subsubsection{Convex Relaxation of (P3)}
Consider an arbitrary SCA iteration $(t+1)$, the reference point is the solution of the relaxed problem in the previous iteration and is denoted as $\{ \hat{\bf f}^{[t]}, c_k^{[t]}, \alpha_{\ell,\ell^{'},m_i}^{[t]} \}$. According to Lemma \ref{Lma:DCProblem}, $R_k(\hat{\bf f})$ and  $Q_{\ell,\ell^{'},m_i}\left( \{c_k \},\alpha_{\ell,\ell^{'},m_i} \right) $ are differntiable and convex, and hence are no less than their corresponding first-order Taylor expansions at the reference point:
\begin{equation}\label{Eq:RQ}
\left\{
\begin{aligned}
& R_k\left(\hat{\bf f}\right) \geq \hat{R}_k^{[t]}\left(\hat{\bf f}\right) , \\
& Q_{\ell,\ell^{'},m_i}\left( \{c_k \},\alpha_{\ell,\ell^{'},m_i} \right) \geq  \hat{Q}_{\ell,\ell^{'},m_i}^{[t]}\left( \{c_k \},\alpha_{\ell,\ell^{'},m_i} \right), \quad \forall (\ell,\ell^{'},m_i).
\end{aligned}
\right.
\end{equation}
In the equation above, $ \hat{R}_k^{[t]}\left(\hat{\bf f}\right)$ and $\hat{Q}_{\ell,\ell^{'},m_i}^{[t]}\left( \{c_k \},\alpha_{\ell,\ell^{'},m_i} \right)$ are the corresponding first-order linear expansion functions, given by 
\begin{equation}\label{Eq:RSubst}
\hat{R}_k^{[t]}\left(\hat{\bf f}\right)  =  R(\hat{\bf f}^{[t]})  + 4 \hat{P}_k(\hat{\bf f} - \hat{\bf f}^{[t]} )^H\left( {\bf h}_k^H  \hat{\bf f}^{[t]} {\bf h}_k \right),\quad 1\leq k \leq K, 
\end{equation}
and
\begin{equation}\label{Eq:QSubst}
\begin{aligned}
\hat{Q}_{\ell,\ell^{'},m_i}^{[t]}\left( \{c_k \},\alpha_{\ell,\ell^{'},m_i} \right) = &Q\left( \{c_k^{[t]} \},\alpha_{\ell,\ell^{'},m_i}^{[t]} \right)+  \sum_{k=1}^{K} A_k^{[t]} \left(  c_k - c_k^{[t]} \right) \\
 & +  B_{\ell,\ell^{'},m_i}^{[t]}  ( \alpha_{\ell,\ell^{'},m_i} - \alpha_{\ell,\ell^{'},m_i}^{[t] }),
\end{aligned}
\end{equation}
where
\begin{equation}
\left\{
\begin{aligned}
&A_k^{[t]} = \dfrac{\partial Q }{ \partial c_k }\bigg|_{c_k =c_k^{[t]}} = \dfrac{ 2 \sum\nolimits_{k=1}^K c_k^{[t]}  \left( \mu_{\ell,m_i} - \mu_{\ell^{'},m_i}\right)^2 }{\alpha_{\ell,\ell^{'},m_i}^{[t]} },\\
&  B_{\ell,\ell^{'},m_i}^{[t]}  = \dfrac{\partial Q }{ \partial  \alpha_{\ell,\ell^{'},m_i} }\bigg|_{ \alpha_{\ell,\ell^{'},m_i} = \alpha_{\ell,\ell^{'},m_i}^{[t]}} =  -   \left[ \dfrac{ \left( \sum\nolimits_{k=1}^K c_k^{[t]} \right) \left( \mu_{\ell,m_i} - \mu_{\ell^{'},m_i}\right) }{ \alpha_{\ell,\ell^{'},m_i}^{[t] } } \right]^2.
\end{aligned}
\right.
\end{equation}

Next, by substituting the inequalities in \eqref{Eq:RQRelaxation} into (P3), a relaxed problem can be derived as
\begin{equation*}(\text{P4})\;
\begin{aligned}
\max\limits_{\{c_k\},\hat{\bf f} ,\{ \alpha_{\ell,\ell^{'},m_i} \}  }&\;\; G =  \dfrac{2}{L(L-1)} \sum\limits_{i=1}^2 \sum\limits_{\ell^{'} =1}^L \sum\limits_{\ell<\ell^{'} } \alpha_{\ell,\ell^{'},m_i},\\
\text{s.t.} \;\;& c_k^2 \leq \hat{R}_k^{[t]}\left(\hat{\bf f}\right)  , \quad 1\leq k \leq K,\\
& \hat{Q}_{\ell,\ell^{'},m_i}^{[t]}\left( \{c_k \},\alpha_{\ell,\ell^{'},m_i} \right) - \left( \sum\limits_{k=1}^K c_k \right)^2  \sigma_{m_i}^2  \geq   \sum\limits_{k=1}^K c_k^2 \epsilon_k^2 +  \delta_0^2 \hat{\bf f}^T \hat{\bf f},\; \forall (\ell,\ell^{'},m_i),
\end{aligned}
\end{equation*}
where $\hat{R}_k^{[t]}\left(\hat{\bf f}\right)$ and $\hat{Q}_{\ell,\ell^{'},m_i}^{[t]}\left( \{c_k \},\alpha_{\ell,\ell^{'},m_i} \right)$ are defined in \eqref{Eq:RSubst} and \eqref{Eq:QSubst}, respectively.  (P4) is convex. The proof is straightforward and hence omitted. To address (P4), the well-known CVX toolbox can be used \cite{grant2009cvx}.

\subsubsection{Task-oriented AirComp Design} 
Based on the convex relaxation approach above, (P3) can be addressed by using the SCA method, which iteratively solves the relaxed convex problem (P4), and updates the reference point using the obtained solution. The detailed procedure is summarized in Algorithm \ref{Alg:SCA}.

\begin{algorithm}
\caption{Joint Power Control and Receive Beamforming for Task-oriented AirComp}\label{Alg:SCA}

1:  {\bf Input:}   Channel gains $\{{\bf h}_k\}$.
  
2:  {\bf Initialize} $t=0$ and $\{ \hat{\bf f}^{[0]}, c_k^{[0]}, \alpha_{\ell,\ell^{'},m_i}^{[0]} \}$, which is in the feasible region of (P3).

3:  {\bf Loop}

4:  \quad $t=t+1$.

5:  \quad Derive (P4), based on the reference point $\{ \hat{\bf f}^{[t-1]}, c_k^{[t-1]}, \alpha_{\ell,\ell^{'},m_i}^{[t-1]} \}$.

5:  \quad Solve (P4) and obtain the optimum as  $\{ \hat{\bf f}^{[t]}, c_k^{[t]}, \alpha_{\ell,\ell^{'},m_i}^{[t]} \}$.

6:  {\bf Until Convergence}

7:  The solution is 
\begin{equation*}
\begin{aligned}
 \hat{\bf f}^*=\hat{\bf f}^{[t]},\quad  \left\{ c_k^* = c_k^{[t]},\; 1 \leq k \leq K \right\}, \quad  \left\{\alpha_{\ell,\ell^{'},m_i}^* =  \alpha_{\ell,\ell^{'},m_i}^{[t]}, \; \forall (\ell,\ell^{'},m_i)\right\}.
\end{aligned}
\end{equation*}

8:  {\bf Output}: $ \hat{\bf f}^*$, $\{c_k^*\}$, and $\{\alpha_{\ell,\ell^{'},m_i}^* \}$.

\end{algorithm}


\subsection{A Property of Transmit Power Control}
As mentioned, the SCA based solution obtained by Algorithm \ref{Alg:SCA} satisfies the KKT conditions \cite{razaviyayn2014successive}. In this part, some of these conditions are used to find the structure of the solved steering power of each device. To begin with, the Lagrangian function of (P3) is given by
\begin{equation}
\begin{aligned}
\mathcal{L}_{\text{P3}}& = - \dfrac{2}{L(L-1)} \sum\limits_{i=1}^2 \sum\limits_{\ell^{'} =1}^L \sum\limits_{\ell<\ell^{'} } \alpha_{\ell,\ell^{'},m_i} + \sum\limits_{k=1}^K \beta_k \left( c_k^2 - 2\hat{P}_k{\bf h}_k^H \hat{\bf f}\hat{\bf f}^T{\bf h}_k \right) \\
  + &\sum\limits_{i=1}^2 \sum\limits_{\ell^{'} =1}^L \sum\limits_{\ell<\ell^{'} } \lambda_{\ell,\ell^{'},m_i} \left[  \sum\limits_{k=1}^K c_k^2 \epsilon_k^2 +  \delta_0^2 \hat{\bf f}^T \hat{\bf f} + \sigma_{m_i}^2\left( \sum\limits_{k=1}^K c_k \right)^2 - Q_{\ell,\ell^{'},m_i}\left( \{c_k \},\alpha_{\ell,\ell^{'},m_i} \right) \right],
\end{aligned}
\end{equation}  
where $\{\beta_k\geq 0 \}$ and $\{ \lambda_{\ell,\ell^{'},m_i} \geq 0 \}$ are Lagrange multipliers.

Some useful KKT conditions are given by 
\begin{equation}\label{Eq:KKTConditions}
\left\{
\begin{aligned}
&\dfrac{\partial \mathcal{L}_{\text{P3}}}{\partial c_k} = 0, \; 1 \leq k \leq K,\\
& c_k^2 \leq  2\hat{P}_k{\bf h}_k^H \hat{\bf f}\hat{\bf f}^T{\bf h}_k,  \;\; 1\leq k \leq K,\\
&\beta_k \left( c_k^2 - 2\hat{P}_k{\bf h}_k^H \hat{\bf f}\hat{\bf f}^T{\bf h}_k \right) = 0, \;\; 1\leq k \leq K.
\end{aligned}
\right.
\end{equation}
From the first condition, we can obtain
\begin{equation}\label{Eq:Betak}
   \beta_k c_k  + \sum\limits_{i=1}^2 \sum\limits_{\ell^{'} =1}^L \sum\limits_{\ell<\ell^{'} } \lambda_{\ell,\ell^{'},m_i} \left[ c_k \epsilon_k^2  + \left(\sum\limits_{k=1}^K c_k \right)\left( \sigma_{m_i}^2 - \dfrac{ \left( \mu_{\ell,m_i} - \mu_{\ell^{'},m_i}\right)^2 }{\alpha_{\ell,\ell^{'},m_i} }\right)   \right] = 0,
\end{equation}
for all $1\leq k \leq K$. By using a normalized steering power as $c_k^{'} = c_k/(\sum\nolimits_{k=1}^K c_k)$ and substituting $\beta_k$ in \eqref{Eq:Betak}, the above condition can be further derived as 
\begin{equation}\label{Eq:NormalizedReceivePower1}
c_k^{'} = \dfrac{  \sum\nolimits_{i=1}^2 \sum\nolimits_{\ell^{'} =1}^L \sum\nolimits_{\ell<\ell^{'} } \lambda_{\ell,\ell^{'},m_i} \left(  ( \mu_{\ell,m_i} - \mu_{\ell^{'},m_i})^2 /\alpha_{\ell,\ell^{'},m_i}  -  \sigma_{m_i}^2 \right)  }{ \beta_k   + \sum\nolimits_{i=1}^2 \sum\nolimits_{\ell^{'} =1}^L \sum\nolimits_{\ell<\ell^{'} } \lambda_{\ell,\ell^{'},m_i}  \epsilon_k^2}, \; 1 \leq k \leq K.
\end{equation}
From the second condition in \eqref{Eq:KKTConditions}, we have
\begin{equation}\label{Eq:NormalizedReceivePower2}
c_k \leq \sqrt{ 2 \hat{P}_k{\bf h}_k^H \hat{\bf f}\hat{\bf f}^T{\bf h}_k },
\end{equation}
where the equality is achieved when $\beta_k\neq 0$ according to the third condition in \eqref{Eq:KKTConditions}. Then, togethering with  \eqref{Eq:NormalizedReceivePower1}, the normalized steering power of device $k$ is given as
\begin{equation}\label{Eq:NormalizedReceivePower}
c_k^{'} = \left\{ 
\begin{aligned}
&  \dfrac{  \sum\limits_{i=1}^2 \sum\limits_{\ell^{'} =1}^L \sum\limits_{\ell<\ell^{'} } \lambda_{\ell,\ell^{'},m_i} \left(  ( \mu_{\ell,m_i} - \mu_{\ell^{'},m_i})^2 /\alpha_{\ell,\ell^{'},m_i}  -  \sigma_{m_i}^2 \right)  }{ \sum\limits_{i=1}^2 \sum\limits_{\ell^{'} =1}^L \sum\limits_{\ell<\ell^{'} } \lambda_{\ell,\ell^{'},m_i}  \epsilon_k^2}, \;\; \text{if } \beta_k = 0,\\
& \dfrac{\sqrt{  2 \hat{P}_k{\bf h}_k^H \hat{\bf f}\hat{\bf f}^T{\bf h}_k }   }{ \sum\limits_{k_1=1,\; k_1\neq k}^K c_{k_1} +   \sqrt{ 2 \hat{P}_k{\bf h}_k^H \hat{\bf f}\hat{\bf f}^T{\bf h}_k }}, \;\; \text{if } \beta_k \neq 0,
\end{aligned}
\right.
\end{equation}
where $c_{k_1}$ with $k_1\neq k$ is irrelevant to the channel gain and sensing noise power of device $k$ according to \eqref{Eq:NormalizedReceivePower1} and \eqref{Eq:NormalizedReceivePower2}. Several observations can be made from \eqref{Eq:NormalizedReceivePower}. If the transmit power or the channel magnitude is large enough, i.e., $\beta_k=0$ and the equality in \eqref{Eq:NormalizedReceivePower2} is not achieved, the normalized steering power of device $k$, say $c_k^{'}$, is inversely proportional to its sensing data noise power $\epsilon_k^2$. Otherwise (i.e., $\beta_k\neq 0$), $c_k^{'}$ is an increasing function of its channel magnitude.


\section{Performance Evaluation}\label{Sect:PerformanceEvaluation}

\subsection{Experiment Setup}

\subsubsection{Communication model} 
In this experiment, a multi-user single-input multiple-output network is considered, where $K$ single-antenna devices are distributed randomly within a circle with a radius of $50$ meters. The multi-antenna AP (edge server) is located at the circle center. For the $k$-th device, the channel gain ${\bf h}_{k}$ is modeled as ${\bf h}_{k} = \left\vert \phi_{k}{\bm \rho}_{k} \right\vert^{2}$, where $\phi_{k}$ and ${\bm \rho}_{k}$ stand for the large-scale and small-scale fading propagation coefficients, respectively. The large-scale propagation coefficient (in dB) is modeled as $[\phi_{k}]_{\text{dB}}=-[\text{PL}_{k}]_{\text{dB}} +  [\zeta_{k}]_{\text{dB}}$, where $[\text{PL}_{k}]_{\text{dB}} = 128.1+37.6\log_{10}\text{dist}_{k}$ ($\text{dist}_{k}$ is the distance in kilometer) is the path loss in dB, and $[\zeta_{k}]_{\text{dB}}$ accounts for the shadowing in dB. In the simulation, $[\zeta_{k}]_{\text{dB}}$ is a Gauss-distributed random variable with mean zero and variance $\sigma^{2}_{\zeta}$. Besides, Rayleigh small-scale fading is assumed, i.e., ${\bm \rho}_{k} \sim \mathcal{CN}(0,{\bf I})$. 

\subsubsection{Inference task} 
A concrete classification task of human motion recognition to identify four distinct human motions, i.e., child walking, child pacing, adult walking, and adult pacing, is considered. The wireless sensing simulator proposed in \cite{Li2021SPAWC} is adopted to generate the datasets for this task.  Using similar settings as \cite{matlab}, the heights of adults and children are assumed to be uniformly distributed in the intervals $[1.6\text{m}, 1.9\text{m}]$ and $[0.9\text{m}, 1.2\text{m}]$, respectively. The speeds of standing, walking, and pacing are set as $0$ m/s, $0.5H$ m/s, and $0.25H$ m/s, respectively, where $H$ is the height value. The heading of the moving human is set to be uniformly distributed in $[–180^{\circ}, 180^{\circ}]$. 

\subsubsection{Inference model}
Two AI models based on SVM and MLP neural networks are used for the inference task. 
The neural network model consists of two hidden layers, each with 80 and 40 neurons. The total number of training data samples is 6400, which are assumed to have no noise corruption during the training of both AI models. The testing dataset includes  1600 noise-corrupted data samples, where the noise power is determined by the three schemes.

Unless specified otherwise, other simulation parameters are stated in Table \ref{tab:sensing-parameters}. All experiments are implemented using Python 3.8 on a Linux server with one NVIDIA\textsuperscript{\textregistered} GeForce\textsuperscript{\textregistered} RTX 3090 GPU 24GB and one Intel\textsuperscript{\textregistered} Xeon\textsuperscript{\textregistered} Gold 5218 CPU. 

\begin{table}[tt]
	\caption{Simulation Parameters }
	\label{tab:sensing-parameters}
	\centering
	\begin{tabular}{l l l l}
		\hline
		\bfseries Parameter & \bfseries Value & \bfseries Parameter &  \bfseries Value \\
		\hline\hline
		Number of ISAC devices, $K$ & $3$  & Channel noise variance, $\delta_{0}^{2}$ & 1 \\
		feature noise variance, $\epsilon_k^2$ & 0.4 & Number of receive antennas, $N_{r}$ & 8\\
		Number of dimension after PCA, $N_{K}$ & 12 &	Number of classes, $L$ & 4 \\
		Training data sizes, $B$ & $6400$ &		Transmit power, $P_{k}$ & 12 mdB \\
		Variance of shadow fading, $\sigma_{\zeta}^{2}$ & 8 dB &  Communication channel noise power, $\delta_{c}^{2}$ & $10^{-11}$ W\\
		\hline
	\end{tabular}
\end{table}

\subsection{Inference Algorithms}

For comparison, we consider three schemes as follows.
\begin{itemize}
	\item \textit{Baseline}: In this scheme, random receive beamformer is first used, and then the transmit precoders are selected to satisfy the constraint in (P1).
	\item \textit{Weighted subspace centroid}: All the parameters are allocated following the AirComp scheme in \cite{Zhu2019IoTJ}, where the design criterion is MMSE and channel equalization of all devices is performed.
	\item \textit{ Joint design of transmit precoding and receive beamforming (our proposal)}: All parameters are set to follow the proposed scheme Algorithm \ref{Alg:SCA}.
\end{itemize}

\subsection{Experimental Results}
This part starts from presenting the relation between the discriminant gain and the corresponding inference accuracy of the two models. Then, the three schemes are compared for both models in terms of the changing number of devices and transmit power. Finally, the influence of feature elements' number on the inference accuracy is shown. 

\begin{figure}[t]
	\centering
	\includegraphics[width=0.485\textwidth]{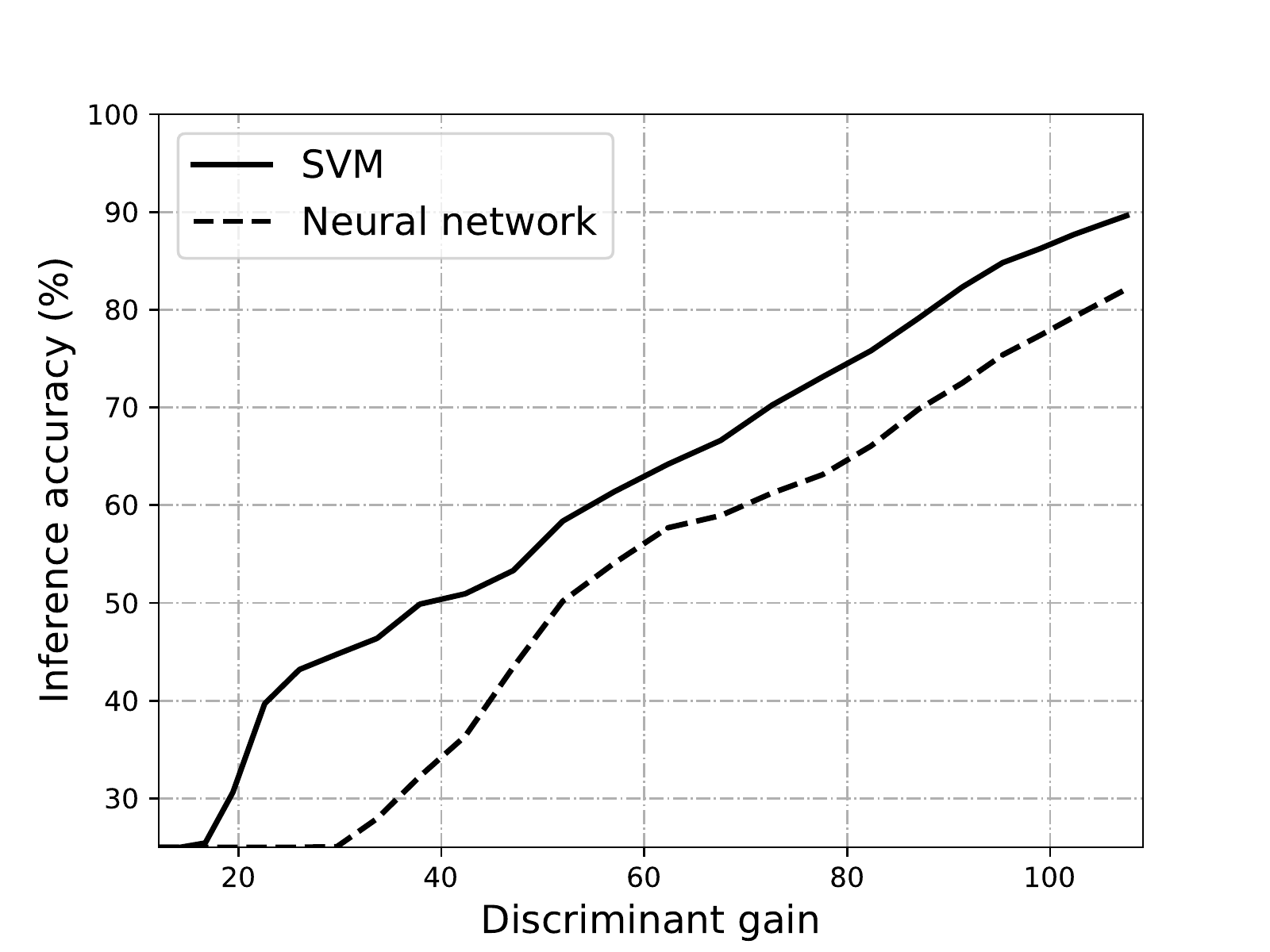}
	\caption{Inference accuracy versus discriminant gain.}
	\vspace{-0.5cm}
	\label{fig:disc_acc}
\end{figure}

\subsubsection{Inference accuracy v.s. discriminant gain}
In Fig. \ref{fig:disc_acc}, the relation between inference accuracy and discriminant gain for the SVM model and the MLP neural network is presented. To investigate the relation, different values of discriminant gain are obtained by using different transmit power on devices. From the figure, for both models, it is seen that the inference accuracy increases as the discriminant gain grows. Additionally, the SVM beats the neural network, as the training of the latter is overfitting, which has a complex model compared to the simple dataset. 

\begin{figure*}[t]
	\centering
	\subfloat[Inference accuracy with SVM versus number of devices]{\includegraphics[width=0.485\textwidth]{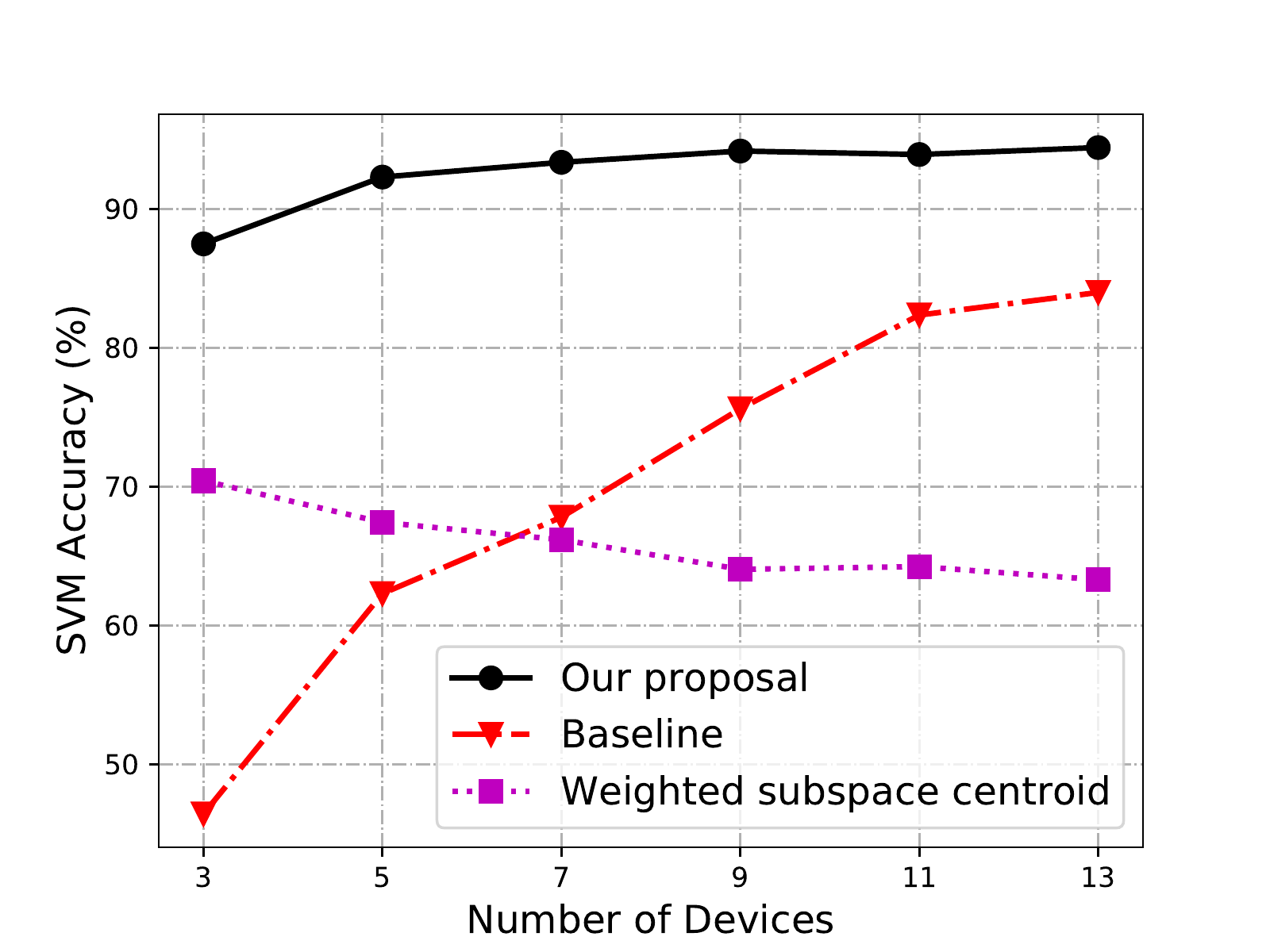}
		\label{fig:svm_device_numbers_accuracy}}\hfil
	\subfloat[Inference accuracy with MLP versus number of devices]{\includegraphics[width=0.485\textwidth]{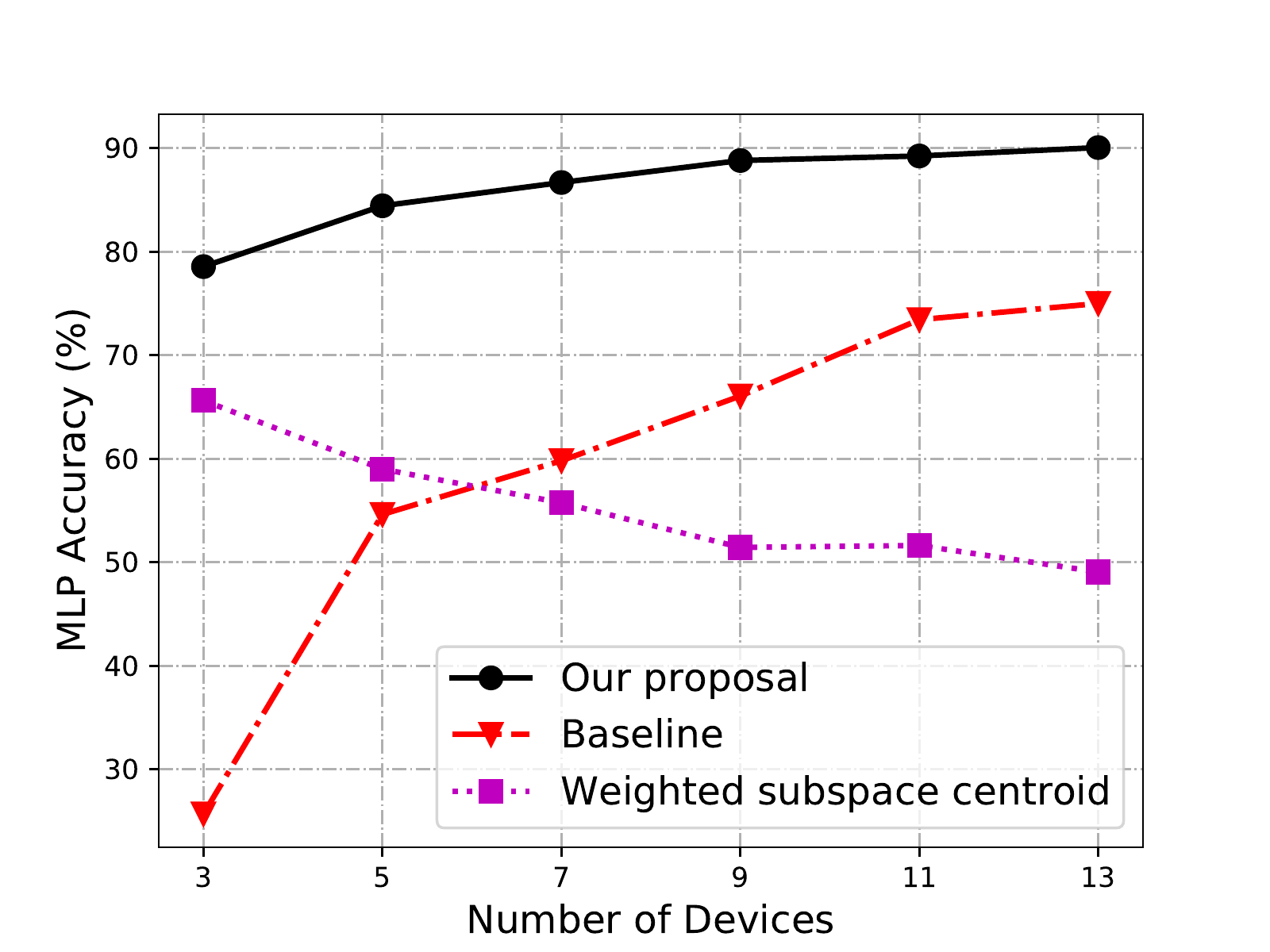}
		\label{fig:mlp_device_numbers_accuracy}}
	\caption{Inference accuracy comparison among different models under differenct number of devices.}
	\vspace{-0.5cm}
	\label{fig:device_number_accuracy}
\end{figure*}

\subsubsection{Inference accuracy v.s. number of devices}
The inference accuracy of both models is shown in Fig. \ref{fig:device_number_accuracy} in terms of a changing number of devices. It is observed that our proposed scheme has the best performance. Besides, the performance of the weighted subspace centroid scheme decreases with the number of devices. The reason is as follows. Channel equalization is performed among all devices under the target of MMSE in this scheme. As a result, with a growing number of devices, the possibility of deep fading channels increases, which leads to a higher distortion level. Better inference accuracy is obtained in the baseline scheme and our proposed scheme, as the number of devices increases. This is because under the task-oriented principle, different steering powers are permitted for different devices, and thus the data diversity provided by more devices can be fully exploited.  
\begin{figure*}[t]
	\centering
	\subfloat[Inference accuracy with SVM versus transmit power]{\includegraphics[width=0.485\textwidth]{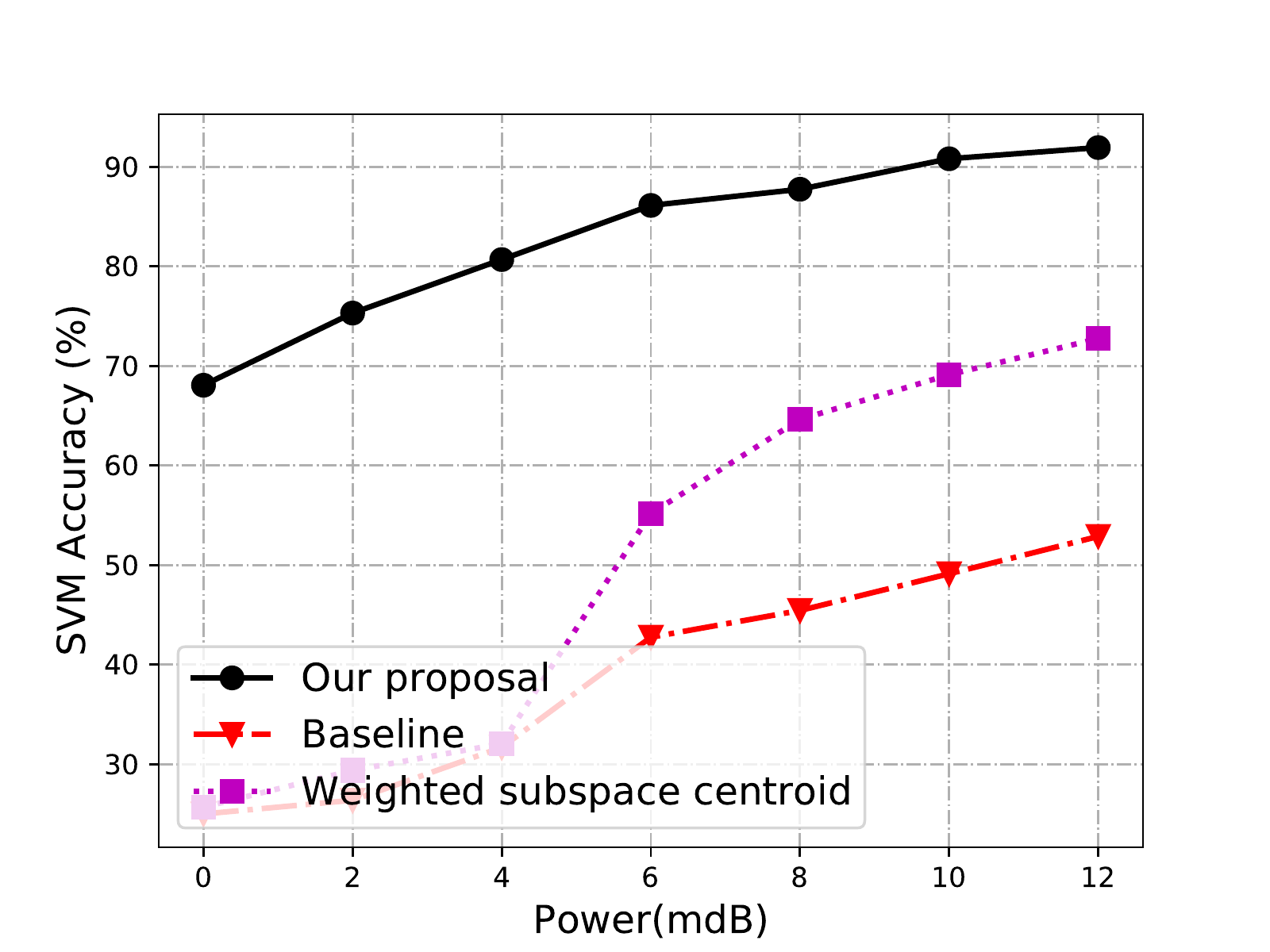}
		\label{fig:svm_power_accuracy}}\hfil
	\subfloat[Inference accuracy with MLP versus transmit power]{\includegraphics[width=0.485\textwidth]{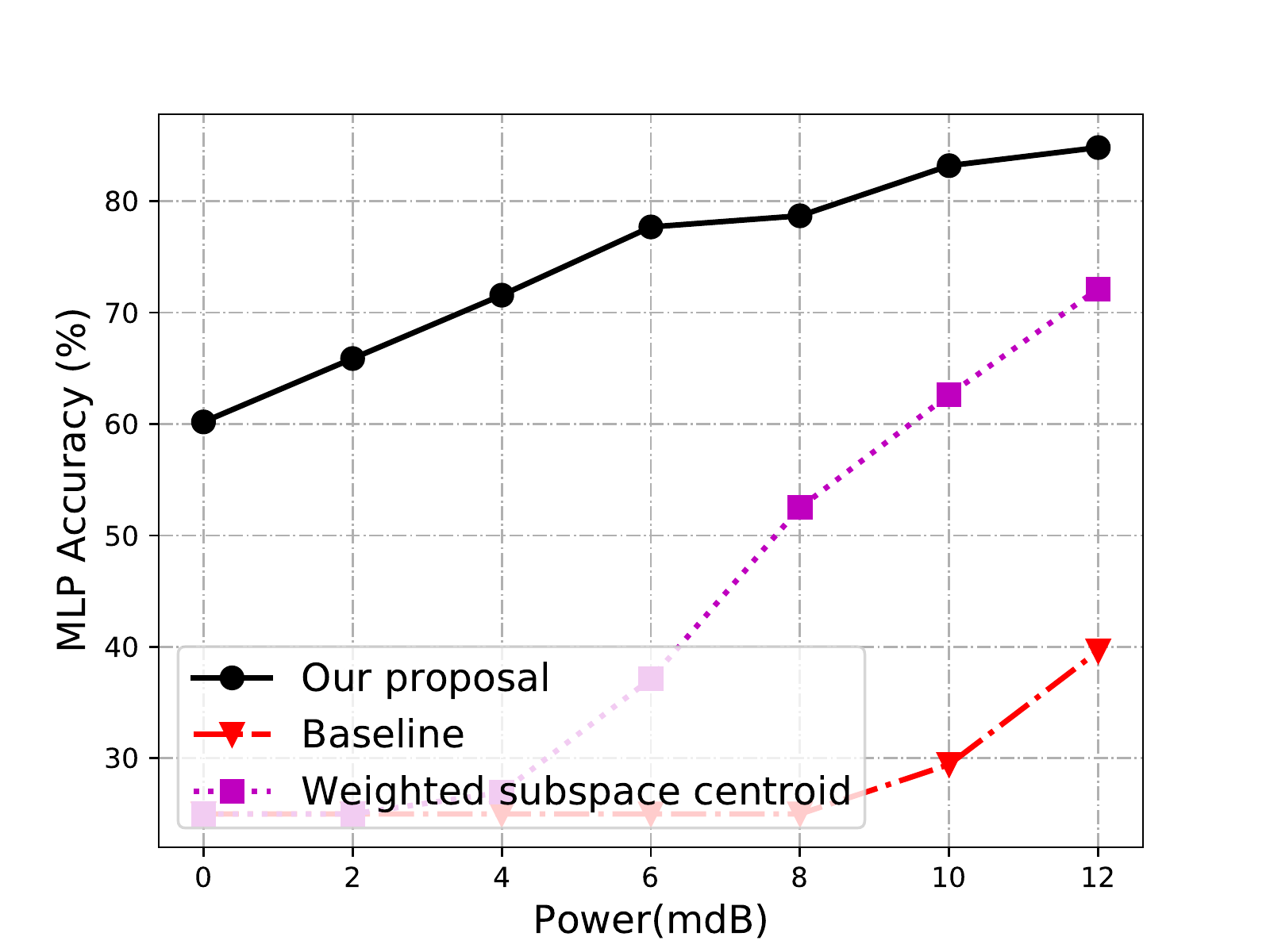}
		\label{fig:mlp_power_accuracy}}
	\caption{Inference accuracy comparison among different models under different transmit power.}
	\vspace{-0.5cm}
	\label{fig:power_accuracy}
\end{figure*}

\subsubsection{Inference accuracy v.s. transmit power}
The inference accuracy of both models under various transmit powers is shown in Fig. \ref{fig:power_accuracy}. In both cases, improved inference accuracy is acquired as the transmit power rises, since larger transmit powers can more effectively suppress the channel noise. As well, our proposed scheme outperforms the other two schemes.

\begin{figure*}[t]
	\centering
	\subfloat[Inference accuracy with SVM versus PCA dimension]{\includegraphics[width=0.485\textwidth]{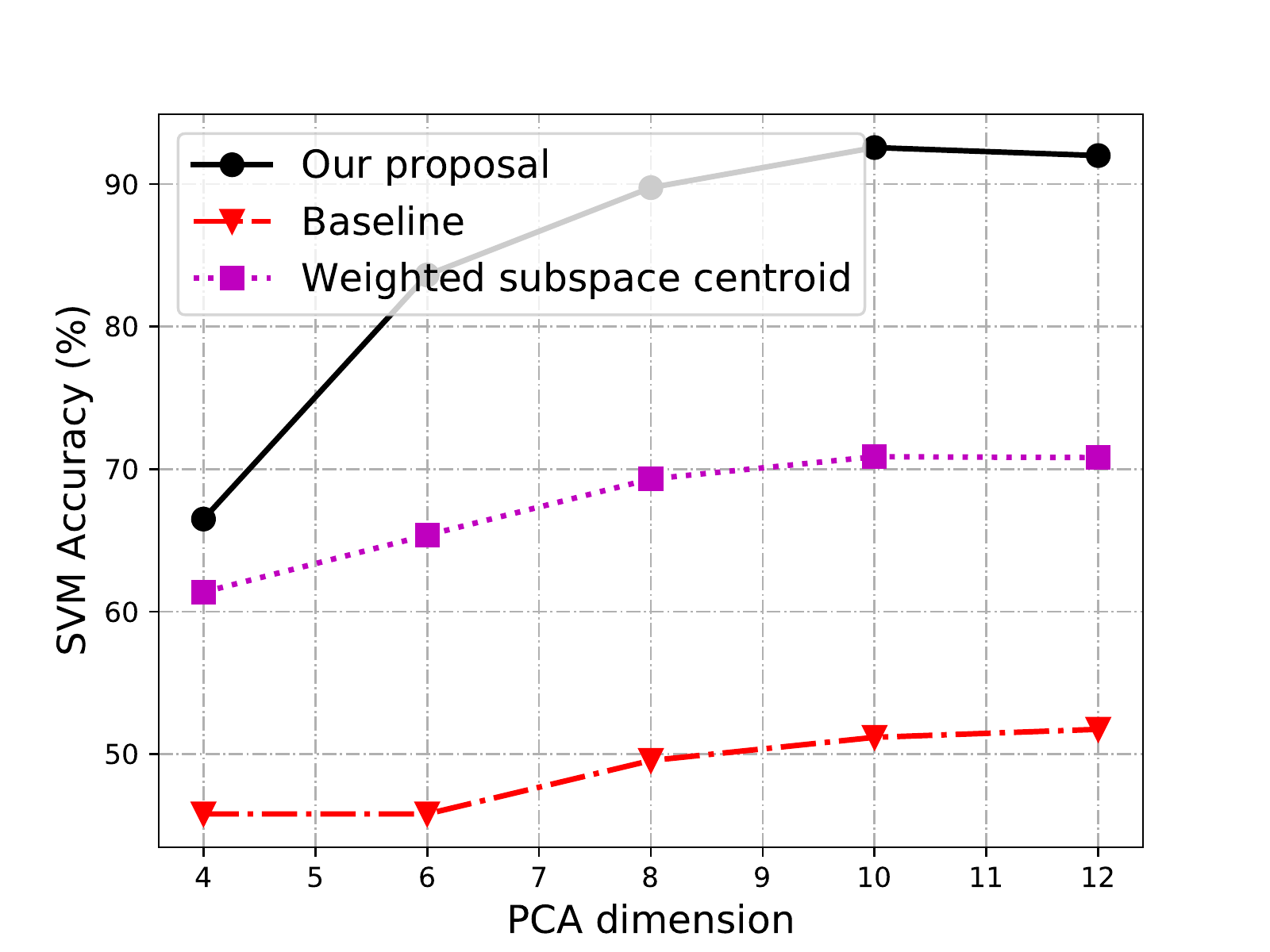}
		\label{fig:svm_PCA_dim_accuracy}}\hfil
	\subfloat[Inference accuracy with MLP versus transmit power]{\includegraphics[width=0.485\textwidth]{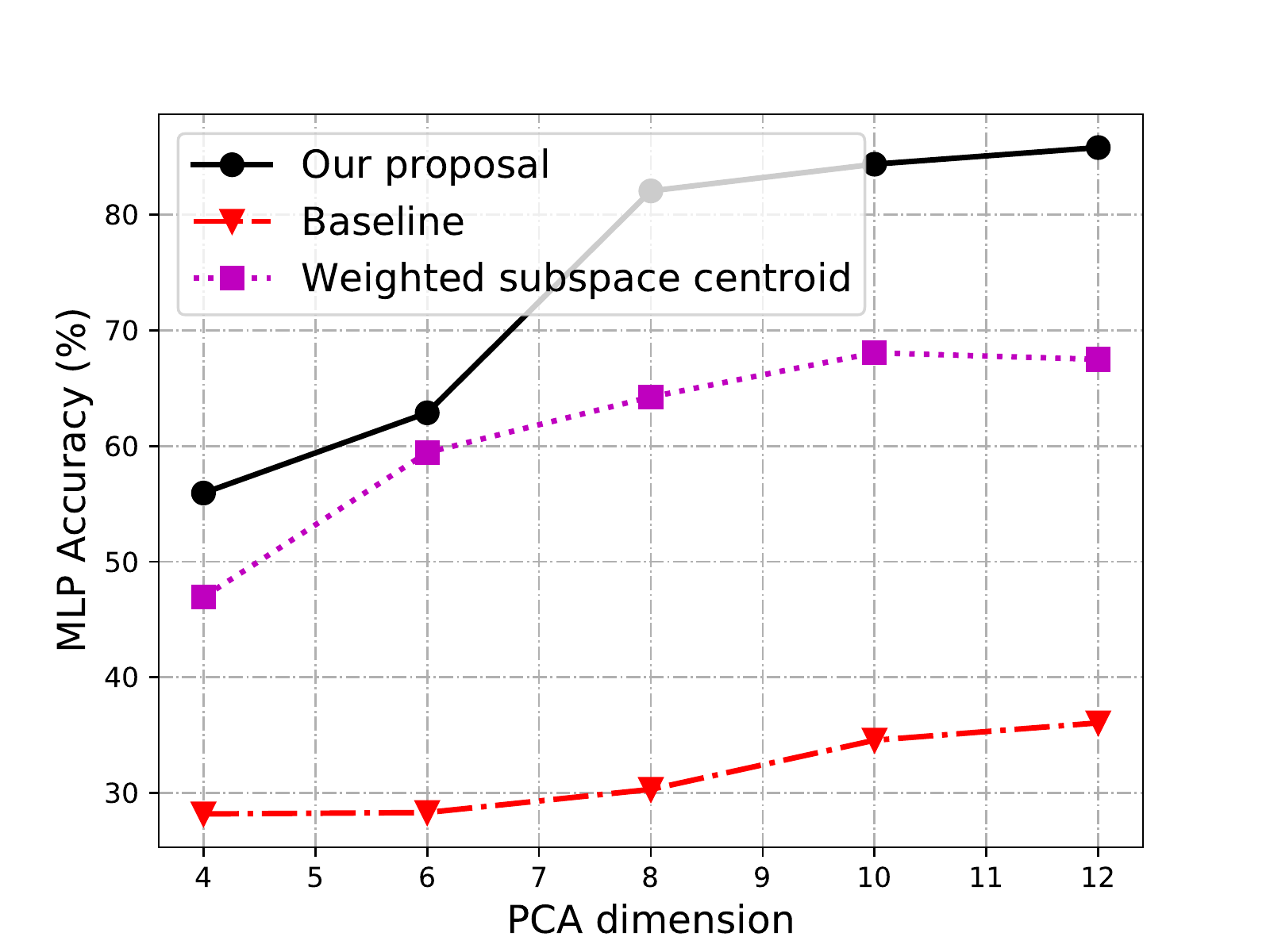}
		\label{fig:mlp_PCA_dim_accuracy}}
	\caption{Inference accuracy comparison among different models under differenct PCA dimension.}
	\vspace{-0.5cm}
	\label{fig:PCA_dimension_accuracy}
\end{figure*}

\subsubsection{Inference accuracy v.s. number of feature elements}
The inference accuracy of both models in terms of the number of used feature elements in the task is shown in Fig \ref{fig:PCA_dimension_accuracy}. Specifically, the number of used feature elements is sequentially increased  following the order of the PCA dimensions from the largest to the least.
From the figure, the inference accuracy increases as the number of used feature elements in the inference task. That's because more feature elements increase the dimensions of feature space so that different classes can be better differentiated and a better discriminant gain can be achieved. In addition, the accuracy turns to be saturate at a large number of feature elements, since  the added least important feature elements have less contribution to the discriminant gain and the inference accuracy.

The extensive experimental results presented above demonstrate the best performance of the proposed optimal scheme and verify our theoretical analysis.

\section{Conclusion}
To enhance the performance of multi-device edge inference systems, this paper proposed a task-oriented AirComp scheme. To alleviate the influence of sensing noise on the inference accuracy, it aggregated the noise-corrupted local feature vectors to generate a global one at the server for completing the task. Instead of using the conventional design criterion MMSE, the task-oriented AirComp scheme aimed at directly maximizing the inference accuracy measured by an approximate but tractable metric, called discriminant gain, which represents the averaged centroids distances of different class pairs normalized by their covariance in the Euclidean feature space. A larger discriminant gain means that the classes are better separated, and thus indicates a higher inference accuracy. This task-oriented problem, however, was non-convex due to the complicated form of discriminant gain as well as the couple of transmit precoding and receive beamforming. To tackle this problem, variables transformation was first applied to derive an equivalent d.c. problem. Then, a joint scheme of transmit precoding and receive beamforming was proposed to address the d.c. problem based on the SCA approach. The performance of the proposed scheme was verified using extensive numerical results of a concrete classification task of human motion recognition.

It is noteworthy that the theoretical analysis presented in this work holds for all types of the AI models, from general linear models like SVM to deep neural networks (DNNs), but highly depends on the assumption that the feature vector follows a mixture of Gaussians distribution. We remark that, in some AI tasks, this assumption may not strictly hold, since either the raw data generated by the source or the feature maps generated by the intermediate layers of a DNN  may not follow Gaussian distribution. To tackle this issue, a practical approach is to first fit the data to a mixture of Gaussians distribution approximating the ground-truth. Then, the proposed scheme can be extended to these AI tasks, and its performance is verified via extensive experiments based on a high-fidelity human-motion recognition dataset (Please refer to Section \ref{Sect:PerformanceEvaluation}).  

This work opens several interesting directions. One is the device selection for further accuracy enhancement by excluding the devices with a weak channel or high sensing noise.  Another is to extend the current design to the case where devices are equipped with multi antennas.

\appendix

\subsection{Proof of Lemma \ref{Lma:LocalDistribution}}\label{Apdx:LmaLocalDistribution}
First, according to \eqref{Eq:ObeservationElement}, $x_{m_i}$ can be decomposed into the avegrage of $L$ indepdent Gaussian variables, as
$x_{m_i} = \dfrac{1}{L} \sum\limits_{\ell=1}^L x_{\ell,m_i}, \quad i=1,2$,
where the distribution of $x_{m_i,\ell}$ is 
\begin{equation}\label{Eq:xlmi}
x_{\ell,m_i} \sim \mathcal{N}\left(\mu_{\ell,m_i}, \sigma_{m_i}^2 \right), \quad 1 \leq \ell \leq L, \; i = 1,2.
\end{equation}
Then, by substituting the above equation into the local elements defined in \eqref{Eq:xmi}, we have 
\begin{equation}
x_{k,m_i} =  \dfrac{1}{L} \sum\limits_{\ell=1}^L x_{\ell,m_i} + d_{k,m_i}, \quad 1 \leq k \leq K,  \; i=1,2.
\end{equation}
It follows that 
\begin{equation}
x_{k,m_i} =  \dfrac{1}{L} \sum\limits_{\ell=1}^L x_{\ell,k,m_i}, \quad 1 \leq k \leq K, \; i=1,2.
\end{equation}
where 
$x_{\ell,k,m_i} = x_{\ell,m_i} + d_{k,m_i}$. 
Next, by substituting  the distributions  of $x_{\ell,m_i}$ in \eqref{Eq:xlmi} and the distribution of $d_{k,m_i}$ in \eqref{Eq:dkmi}, the distribution of $x_{\ell,k,m_i}$ can be derived as 
\begin{equation}
x_{\ell,k,m_i} \sim \mathcal{N}\left(\mu_{\ell,m_i}, \sigma_{m_i}^2 + \epsilon_k^2 \right),  \quad 1 \leq k \leq K, \; \& \; \leq \ell \leq L, \; \& \; i=1,2.
\end{equation}
It follows that the distribution of  $x_{k,m_i}$ can be derived as
\begin{equation}
x_{k,m_i} \sim \dfrac{1}{L}\sum\limits_{\ell =1 }^L \mathcal{N}\left(\mu_{\ell,m_i}, \sigma_{m_i}^2 + \epsilon_k^2 \right), \quad i=1,2,\; \&\; 1\leq k \leq K.
\end{equation}

\subsection{Proof of Lemma \ref{Lma:EstimateDistribution}}\label{Apdx:LmaEstimateDistribution}
First, for the received symbol $\hat{s}$ in \eqref{Eq:EstimateSymbol}, the received noise can be derived as 
\begin{equation}
\begin{aligned}
{\bf f}^H {\bf n} & = ({\bf f}_1+j {\bf f}_2)^H ({\bf n}_1+ j{\bf n}_2)=  {\bf f}_1^T {\bf n}_1 + {\bf f}_2^T {\bf n}_2 + j ( {\bf f}_1^T {\bf n}_2 - {\bf f}_2^T {\bf n}_1 ), 
\end{aligned} 
\end{equation}
where ${\bf f}_1$ and ${\bf f}_2$ are the real part and imaginary part of ${\bf f}$ respectively, and ${\bf n}_1$ and ${\bf n}_2$ are the real part and imaginary part of the Guassian noise ${\bf n}$ respectively. Specifically, we have
\begin{equation}
\begin{aligned}
{\bf n}_1 \sim \mathcal{N}\left({\bf 0}, \dfrac{\delta_0^2}{2} {\bf I} \right), \quad {\bf n}_2 \sim \mathcal{N}\left({\bf 0}, \dfrac{\delta_0^2}{2} {\bf I} \right)
\end{aligned}
\end{equation}
where $\delta_0^2$ is the noise variance. Then, for the real part of the received noise, its expectation can co-variance can be derived as 
\begin{equation}
\mathbb{E}\left[ \mathsf{Re}({\bf f}^H {\bf n}) \right] = \mathbb{E}\left[  {\bf f}_1^T {\bf n}_1 + {\bf f}_2^T {\bf n}_2 \right]  = {\bf 0},
\end{equation}
and 
\begin{equation}
\begin{aligned}
\mathbb{C} = \left[ \mathsf{Re}({\bf f}^H {\bf n}) \right] & = \mathbb{E}\left[  \left( {\bf f}_1^T {\bf n}_1 + {\bf f}_2^T {\bf n}_2  \right)  \left( {\bf f}_1^T {\bf n}_1 + {\bf f}_2^T {\bf n}_2  \right)^T \right] = \dfrac{\delta_0^2}{2}\left(  {\bf f}_1^T {\bf f}_1 + {\bf f}_2^T {\bf f}_2 \right). 
\end{aligned}
\end{equation}
That's to say,
\begin{equation}\label{Eq:RealNoiseDistribution}
 \mathsf{Re}({\bf f}^H {\bf n})  \sim \mathcal{N} \left( {\bf 0},  \dfrac{\delta_0^2}{2}\left(  {\bf f}_1^T {\bf f}_1 + {\bf f}_2^T {\bf f}_2 \right) \right).
\end{equation}
Similarly, it can be derived that the imaginary part of the received noise has the same distribution:
\begin{equation}\label{Eq:ImgNoiseDistribution}
 \mathsf{Im}({\bf f}^H {\bf n})  \sim \mathcal{N} \left( {\bf 0},  \dfrac{\delta_0^2}{2}\left(  {\bf f}_1^T {\bf f}_1 + {\bf f}_2^T {\bf f}_2 \right) \right).
\end{equation}
By substituting the noise distributions in \eqref{Eq:RealNoiseDistribution} and  \eqref{Eq:ImgNoiseDistribution} into the global estimates in \eqref{Eq:Estimate} and using the similar method in Appendix \ref{Apdx:LmaLocalDistribution}, i.e., decomposing the local estimate $x_{k,m_i}$ into the average of $L$ independent Gaussia variables, the distributions of the  global estimates can be derived as in \eqref{Eq:EstimateDistribution}.

\subsection{Proof of Lemma \ref{Lma:SymmteicBeamformers}}\label{Apdx:LmaSymmteicBeamformers}
The Lagrange function of the problem in \eqref{Eq:MP1} can be written as
\begin{equation}
\begin{aligned}
\mathcal{L} = & -\dfrac{2}{L(L-1)} \sum\limits_{i=1}^2 \sum\limits_{\ell^{'} =1}^L \sum\limits_{\ell<\ell^{'} } \alpha_{\ell,\ell^{'},m_i} + \sum\limits_{k=1}^K \beta_k \left[ c_k^2 - \hat{P}_k  {\bf h}_k^H \left( {\bf f}_1{\bf f}_1^T + {\bf f}_2{\bf f}_2^T \right) {\bf h}_k \right] \\
&+ \sum\limits_{i=1}^2 \sum\limits_{\ell^{'} =1}^L \sum\limits_{\ell<\ell^{'} } \lambda_{\ell,\ell^{'},m_i} \left[  \alpha_{\ell,\ell^{'},m_i} \hat{\sigma}_{m_i}^2 - \left(\hat{\mu}_{\ell,m_i} - \hat{\mu}_{\ell^{'},m_i}\right)^2 \right],
\end{aligned}
\end{equation}
where  $\left(\hat{\mu}_{\ell,m_i} - \hat{\mu}_{\ell^{'},m_i}\right)^2$ is defined in \eqref{Eq:MuDifference} and $\hat{\sigma}_{m_i}^2$ is defined in \eqref{Eq:Sigma}. KKT conditions are necessary to achieve the optimal solution. Some useful KKT conditions are given below.
\begin{equation}
\begin{aligned}
&\dfrac{ \partial \mathcal{L} }{ \partial {\bf f}_1 } =  - 2\sum\limits_{k=1}^K \beta_k \hat{P}_k  {\bf h}_k^H {\bf h}_k {\bf f}_1 + \sum\limits_{i=1}^2 \sum\limits_{\ell^{'} =1}^L \sum\limits_{\ell<\ell^{'} } \lambda_{\ell,\ell^{'},m_i} \alpha_{\ell,\ell^{'},m_i} \delta_0^2 {\bf f}_1 = 0,\\
&\dfrac{ \partial \mathcal{L} }{ \partial {\bf f}_2 } =  - 2\sum\limits_{k=1}^K \beta_k P_k  {\bf h}_k^H {\bf h}_k {\bf f}_2 + \sum\limits_{i=1}^2 \sum\limits_{\ell^{'} =1}^L \sum\limits_{\ell<\ell^{'} } \lambda_{\ell,\ell^{'},m_i}  \alpha_{\ell,\ell^{'},m_i} \delta_0^2 {\bf f}_2 = 0.
\end{aligned}
\end{equation}
It is observed from the above equations that ${\bf f}_1 = {\bf f}_2$ won't influence the optimality of the problem. 

\subsection{Proof of Lemma \ref{Lma:ExtendedRegion}}\label{Apdx:LmaExtendedRegion}
First, the second constraint in \eqref{Eq:MP1} can be equally written as
\begin{equation}\label{Eq:Apdx1}
\left(\hat{\mu}_{\ell,m_i} - \hat{\mu}_{\ell^{'},m_i}\right)^2 \geq \alpha_{\ell,\ell^{'},m_i} \hat{\sigma}_{m_i}^2 ,\quad \forall (\ell,\ell^{'},m_i),
\end{equation}
The reason is as follows. In \eqref{Eq:Apdx1}, if the equality is not achieved, the value of $ \alpha_{\ell,\ell^{'},m_i}$ can be increased to make the objective function in \eqref{Eq:MP1} larger. In other words, it's necessary to achieve equality for obtaining the optimal solution. Then, by substituting  $\left(\hat{\mu}_{\ell,m_i} - \hat{\mu}_{\ell^{'},m_i}\right)^2$ in \eqref{Eq:MuDifference} and $\hat{\sigma}_{m_i}^2$  in \eqref{Eq:Sigma} into \eqref{Eq:Apdx1}, it can be derived as
\begin{equation*}
\left( \sum\limits_{k=1}^K c_k\right)^2 \left( \mu_{\ell,m_i} - \mu_{\ell^{'},m_i}\right)^2 \geq  \alpha_{\ell,\ell^{'},m_i} \left[  \sigma_{m_i}^2 \left(  \sum\limits_{k=1}^K c_k \right)^2+ \sum\limits_{k=1}^K c_k^2 \epsilon_k^2 + \delta_0^2 \hat{\bf f}^T \hat{\bf f} \right],\; \forall (\ell,\ell^{'},m_i),
\end{equation*}
It follows that 
\begin{equation}\label{Eq:Apdx2}
\left( \sum\limits_{k=1}^K c_k \right)^2  \left[\dfrac{ \left( \mu_{\ell,m_i} - \mu_{\ell^{'},m_i}\right)^2 }{\alpha_{\ell,\ell^{'},m_i} } - \sigma_{m_i}^2 \right] \geq   \sum\limits_{k=1}^K c_k^2 \epsilon_k^2 +  \delta_0^2 \hat{\bf f}^T \hat{\bf f} , \quad \forall (\ell,\ell^{'},m_i).
\end{equation}

\subsection{Proof of Lemma \ref{Lma:DCProblem} }\label{Apdx:LmaDCProblem}
It is straightforward that the objective function, $ c_k^2$, $R_k(\hat{\bf f}) $, and $ \sum\limits_{k=1}^K c_k^2 \epsilon_k^2 +  \delta_0^2 \hat{\bf f}^T \hat{\bf f} +\sigma_{m_i}^2\left( \sum\limits_{k=1}^K c_k \right)^2$ are convex and differentiable, since they are either linear or combination of quadratic functions. In the sequel, we show that $Q_{\ell,\ell^{'},m_i}\left( \{c_k \},\alpha_{\ell,\ell^{'},m_i} \right)$, are convex  and differentiable. To begin with, $Q\left( \{c_k \},\alpha_{\ell,\ell^{'},m_i} \right)$ can be linearly transformed from the convex function, say $f(x,y) = \dfrac{x^2}{y}, \; x>0, \; y>0$.
As linear transformation preserves convexity,  $Q\left( \{c_k \},\alpha_{\ell,\ell^{'},m_i} \right)$ is convex and differentiable. 

\bibliographystyle{IEEEtran}
\bibliography{reference}

\end{document}